\documentclass[reprint,superscriptaddress,amsmath,amssymb,aps,prd,nofootinbin,floatfix]{revtex4-1}

\usepackage{hyperref}
\usepackage{algorithm}
\usepackage{algorithmic}
\usepackage{amsmath}
\usepackage{multirow}
\usepackage{silence}

\usepackage{graphicx}
\usepackage{dcolumn}
\usepackage{bm}
\newcommand{\apjs}{ApJ}
\newcommand{\procspie}{PROCSPIE}
\newcommand{\mnras}{MNRAS}
\newcommand{\physrep}{PhysRep}
\newcommand{\jcap}{JCAP}

\begin{document}

\preprint{APS/123-QED}

\title{Removal of Galactic foregrounds for the Simons Observatory primordial gravitational wave search}

\author{Ben Thorne}
\affiliation{University of Oxford, Denys Wilkinson Building, Keble Road, Oxford OX1 3RH, UK}
\affiliation{Department of Astrophysical Sciences, Peyton Hall, Princeton University, Princeton, NJ, USA 08540}
\author{Jo Dunkley}%
\affiliation{Department of Astrophysical Sciences, Peyton Hall, Princeton University, Princeton, NJ, USA 08540}
\author{David Alonso}
\affiliation{University of Oxford, Denys Wilkinson Building, Keble Road, Oxford OX1 3RH, UK}
\author{Maximilian H. Abitbol}
\affiliation{University of Oxford, Denys Wilkinson Building, Keble Road, Oxford OX1 3RH, UK}
\author{Josquin Errard}
\affiliation{AstroParticule et Cosmologie, Univ Paris Diderot, CNRS/IN2P3,CEA/Irfu, Obs de Paris, Sorbonne Paris Cit\'{e}, France}
\author{J. Colin Hill}
\affiliation{School of Natural Sciences, Institute for Advanced Study, Princeton, NJ, USA 08540}
\affiliation{Center for Computational Astrophysics, Flatiron Institute, New York, NY, USA 10003}
\author{Brian Keating}
\affiliation{Department of Physics, University of California San Diego, CA, 92093 USA}
\author{Grant Teply}
\affiliation{Department of Physics, University of California San Diego, CA, 92093 USA}
\author{Edward J. Wollack}
\affiliation{NASA Goddard Space Flight Center, 8800 Greenbelt Road, Greenbelt, MD 20771}

\date{\today}

\begin{abstract}
  Upcoming observations from the Simons Observatory have been projected to
  constrain the tensor-to-scalar ratio, $r$, at the level of
  $\sigma(r)=$0.003 \citep{so_forecasts:2018, hui/etal:2018}. Here we describe one of the
  forecasting algorithms for the Simons Observatory in more detail, based on cleaning CMB polarization
  maps using a parametric model. We present a new code to perform this
  end-to-end forecast, and explore the assumptions in greater detail. If
  spatial uniformity of the spectral energy distribution of synchrotron
  radiation and thermal dust emission is assumed over the region planned for
  observations, covering almost a fifth of the sky, a bias of order 1--3$\sigma$
  in $r$ is projected for foreground models consistent with current data.
  We find that by masking the most contaminated regions of sky, or by adopting
  more parameters to describe the spatial variation in spectral index for
  synchrotron and dust, such a bias can be mitigated for the foreground models
  we consider. We also explore strategies for testing whether the cleaned CMB
  polarization maps contain residual foreground contamination, including
  cross-correlating with maps tracing the foregrounds. This method also has
  applications for other CMB polarization experiments.
\end{abstract}

\maketitle


\section{Introduction} \label{sec:intro}
Over the past two decades the anisotropies in the temperature of the Cosmic
Microwave Background (CMB) have been measured with increasing accuracy. These
measurements  are compatible with the $\Lambda$CDM cosmological model, and
have allowed us to constrain its parameters with outstanding accuracy
\citep[e.g.,][]{planck_cosmo:2018}.

Additional information about the history of the Universe is contained in the
polarized anisotropies of the CMB. The CMB is linearly polarized, and can be
decomposed into two (pseudo)-scalar fields, $E$ and $B$. Primary $E$ modes are
sourced by scalar perturbations, but primary $B$ modes are only sourced by
tensor perturbations produced by gravitational waves, making them a target for
constraining inflationary models
\citep[e.g.,][]{kamkionkowski/etal:1997,seljak/zaldarriaga:1996} or testing
alternative models for the early universe \citep{ijjas/steinhardt:2018}.
Measurements of primordial $B$ modes are usually parameterized by the
tensor-to-scalar ratio, $r$, defined to be the ratio between the power in
tensor perturbations and scalar perturbations at a specific scale. Currently,
the best constraints on $r$ ($r_{0.05~{{\rm Mpc}^{-1}}} < 0.07, 95\%~{\rm C.L.}$)
come from BICEP2 / Keck Array data combined with Planck and WMAP data
\citep{bicep2/keck:2018}.

Many experiments, including ACT \citep{actpol_instrument:2016}, SPT-3G
\citep{sptpol3g}, BICEP3 / Keck array \citep{bicep3:2017}, the Simons Array
\citep{polarbear2/simonsarray:2016} and CLASS \cite{chuss/etal:2016} are
measuring the polarized anisotropies of the CMB with improved precision, over a
range of scales. The Simons Observatory \citep{so_forecasts:2018} is an
experiment for the 2020s that will consist of multiple telescopes in the
Chilean Atacama desert. It will have a 6~m  Large Aperture Telescope
(LAT) with arcminute-scale resolution, and an array of three
0.42~m refracting Small Aperture Telescopes (SAT). It is these that are targeted
at measuring degree scale $B$ modes. During a similar time frame the BICEP
Array will operate from the South Pole, also targeting degree-scale B-modes
\cite{hui/etal:2018}.

Other processes on the sky produce $B$ modes, which may contaminate observations
of CMB primary $B$ modes, leading to a bias in the estimation of $r$, or an
increase in the uncertainty, $\sigma(r)$. Polarized radiation from sources within
our own Galaxy is well-known to contaminate large scale observations in all
directions and at all frequencies \citep[e.g.,][]{krachmalnicoff/etal:2018}.
On sub-degree scales, gravitational lensing of the CMB by structure between the
surface of large scattering and today mixes primary $E$ and $B$ modes
\citep{lewis/challinor:2006}, acting as an additional source of confusion noise
for primordial $B$ modes.

As the sensitivity of $B$ mode observations improves, our ability to make
inferences about cosmology will be limited by our modeling of both
gravitational lensing and polarized Galactic foregrounds. To date, an array of
methods for modeling Galactic foregrounds have been suggested
\citep[e.g,][]{ eriksen/etal:2008, alonso17_simul_forec_primor_b_searc, core_component_separation:2018}

In this paper we present a foreground-removal pipeline similar to that used in
\cite{alonso17_simul_forec_primor_b_searc}, which fits the foregrounds
parametrically. We use it to demonstrate the ability of Simons Observatory to
place improved constraints on the tensor-to-scalar ratio in the presence of
large-scale Galactic foregrounds. It is a new implementation of one of the
algorithms already described in \citet{so_forecasts:2018}, hereafter SO19.  We
find that the nominal design of SO, with simple foregrounds, no delensing, and
in the absence of additional systematic uncertainties, should achieve a
constraint of $\sigma(r) \le 0.003$. In this paper we extend the forecasts
presented in SO19, exploring the effect of masking and fitting the foregrounds
with spatially varying spectral parameters. This method should also be applicable
for analysis of data from the BICEP Array and other CMB polarization experiments.

We also also show how cross-correlations of the cleaned maps with Galactic
tracers can be used to detect residual foregrounds.

The paper is structured as follows: in Section \S \ref{sec:simulations} we
describe the synthetic observations, in \S \ref{sec:component_separation} the
component separation technique used, the estimation of power spectra from
cleaned CMB maps, and the inference of the tensor-to-scalar ratio. In
\S \ref{sec:results} we present the results of applying our pipeline to the SO
design, and in \S \ref{sec:conclusions} discuss the results and their
implications.

\section{Simulations}

\label{sec:simulations}
In this section we describe the simulations on which these forecasts are based.
As in SO19 we use the
{\tt PySM}\footnote{\url{https://github.com/bthorne93/PySM_public}} software
\citep{thorne17_python_sky_model} to produce $Q$ and $U$ maps of Galactic dust
and synchrotron emission at the \(n_{\rm freqs}\) frequencies observed by SO.
Using bold-face font to represent vectors, and sans-serif font to indicate
matrix quantities, {\tt PySM} models may be summarized by:
\begin{equation}
  \mathbf{s}(\hat n) = {\sf F}({\bm \beta}(\hat n)) \cdot {\bf T}(\hat n),
\end{equation}
where $\hat n$ is a unit vector in the direction $(\theta,~\phi)$,
 \(\mathbf{s}(\hat n)\) is a \(n_{{\rm freq}} \times  n_{\rm pol}
\times n_{\rm pix}\) vector containing $n_{\rm pol}$ maps with $n_{\rm pix}$
pixels, ${\bf T}$ is a \(n_{\rm comp} \times n_{\rm pol} \times n_{\rm pix}\)
vector containing templates of the emission of each of the \(n_{\rm comp}\)
components at a frequency $\nu_0$, and ${\sf F}$ is a \((n_{{\rm freq}} \times
n_{\rm pol} \times n_{\rm pix}) \times (n_{\rm comp} \times n_{\rm pol} \times
n_{\rm pix}) \) matrix, containing the component SEDs that scale each component
from its reference frequency to the observed frequency $\nu$, and ${\bm \beta}$
represents the parameters of the assumed model SED, which may be spatially
varying.

We follow SO19 by modeling instrument beam, ${\bf B}(\hat n)$, as a symmetric
Gaussian, parameterized by a full width at half maximum,
\(\theta^{\rm FWHM}_{\nu}\), and non-uniform correlated noise,
\(\mathbf{n}(\hat n)\):
\begin{equation}
  \label{eq:sim_eq}
  \mathbf{d}(\hat n) = {\bf s}(\hat n) \circledast {\bf B}(\hat n) + {\bf n}
  (\hat n).
\end{equation}
We use the synthetic observations \({\bf d}(\hat n)\) as inputs to the component
separation algorithm. In the rest of this section we elaborate on the model
choices made in each of the simulation steps.

\subsection{Galactic simulations} \label{sec:galactic_sky}

We consider two foreground models in this study, both of which have polarized
synchrotron and dust components. The synchrotron emission is produced by
fast-moving electrons interacting with the Galactic magnetic field. Infrared
emission from dust grains in the interstellar medium comes from their absorbtion
of light from the interstellar radiation field. Both components are polarized due
to alignment by the Galactic magnetic field.

\begin{figure}[t]
  \centering
  \includegraphics[]{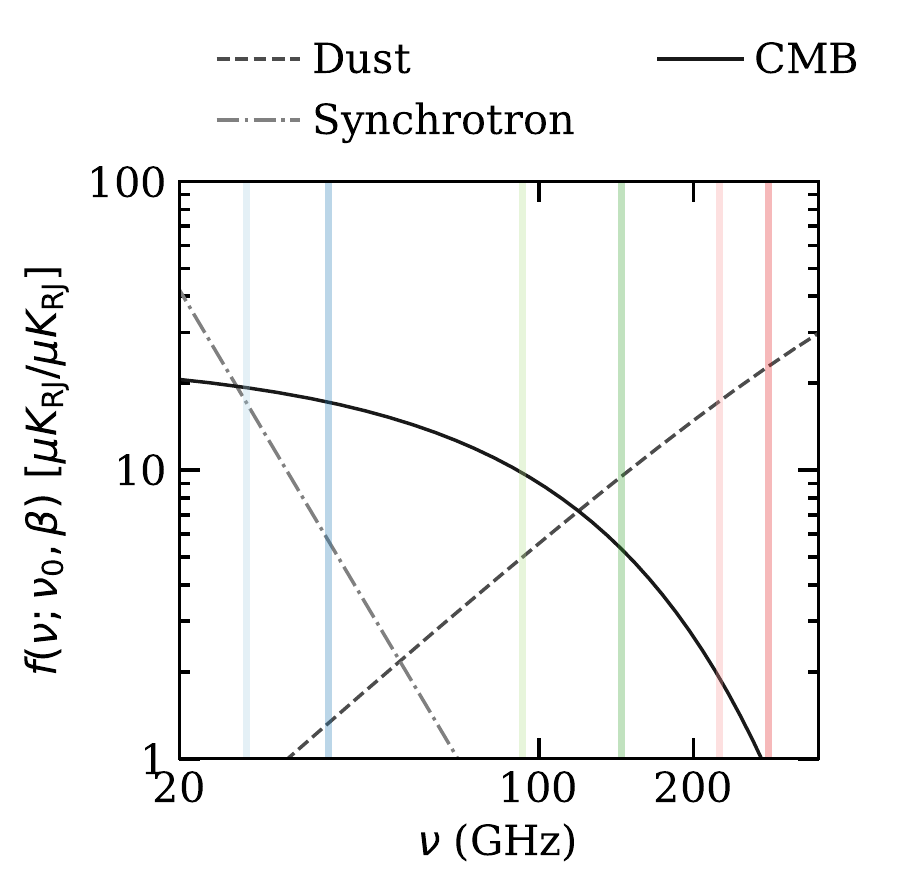}
  \caption{The analytic model SEDs of the CMB, thermal dust, and synchrotron
    described in Section~\ref{sec:galactic_sky}. Amplitudes have been rescaled
    to compare just the shapes of the curves. The vertical lines indicate
    frequencies at which SO is due to make observations; two channels
    characterize
    the low frequency synchrotron, two channels characterize the high frequency
    dust, and two channels observe the CMB around $100$ GHz.}
    \label{fig:sed_comparison}
\end{figure}

The two {\tt PySM} models we consider are:\\

{\sc simset1} This corresponds to the `a1d1f1s1' model of {\tt PySM}, and the
`{\it standard}' model of SO19. It has power-law synchrotron emission with a
spatially varying synchrotron index estimated from WMAP data
\cite{bennett13_nine_yearw_microw_anisot_probe_wmap_obser}.  The dust emission
is described by a modified blackbody with $ I_\nu \propto \nu^\beta B_{\nu}(T)$,
where \(T\) is the temperature of the dust, and \(\beta\) is the opacity index.
It has a spatially varying dust emissivity and temperature estimated from
Planck data. The emission of each component as a function of frequency is
shown in Fig \ref{fig:sed_comparison}.\\

{\sc simset2} This model modifies the synchrotron spectral index map of
{\sc simset1} by adding power at small scales using a Gaussian realization of a
power law power spectrum $\propto \ell^{-2.6}$ \citep{krachmalnicoff/etal:2018}.
This  model is referred to as the `{\it high-res} $\beta_s$' model used in SO19.

\subsection{CMB}

The {\tt PySM} code simulates the primary CMB by creating Gaussian realizations of a
given set of theoretical power spectra, calculated for a given cosmology. We use
a theoretical power spectrum for the fiducial Planck 2018 cosmological
parameters \citep{planck_cosmo:2018}, with no tensor-to-scalar ratio, $r=0$.
{\tt PySM} uses the {\tt healpy} implementation of {\sc synfast} to generate
primary CMB realizations of temperature and polarization.
It then uses the {\tt taylens}\footnote{\url{https://github.com/amaurea/taylens}}
\citep{naess/louis:2013} software to apply the displacements of the primary CMB
caused by gravitational lensing.

\begin{table*}[t]
  \renewcommand{\thetable}{\arabic{table}}
  \centering
  \caption{Simons Observatory expected instrument properties, from
  	\citet{so_forecasts:2018}. The noise levels $\sigma_{I}$ are the intensity
  	white noise levels in $\mu K {\rm amin}$ for a sky area of
  	$f_{\rm sky}=0.1$, with polarization noise $\sqrt{2}$ higher. The parameters
  	$\ell_{\rm knee}$ and  $\alpha_{\rm knee}$ quantify the \(1/f\) model in
  	Eq.~\ref{eq:instrument_noise}. Here $\theta_{\rm FWHM}$ is the full-width at
    half-maximum in arcminutes which in reality varies as a function of frequency.
    In SO19 it was found that the low resolution of the 27 GHz and 39 GHz channels
    does not limit the constraints on $r$. Therefore, for simplicity we use
    $\theta_{\rm FWHM}= 30^\prime$ for all channels.
  }
  \label{tab:somatrix}
  \begin{tabular}{c|c|c|c|c|c|c}
    \hline
    \hline
    \multirow{2}{*}{Frequency (GHz)} & \multicolumn{2}{c|}{$\ell_{{\rm knee}}$} & \multirow{2}{*}{$\alpha_{{\rm knee}}$}& \multicolumn{2}{c|}{\(\sigma_I\) (\(\mu {\rm K \ amin}\))} & \multirow{2}{*}{\(\theta_{\rm FWHM} \ (^\prime)\)} \\
    \cline{2-3} \cline{5-6}    & optimistic &  pessimistic & &  goal & baseline &  \\
    \hline
    \hline
    27  & 30  & 15     & -2.4 & 25  &  35        &  91 \\
    39  & 30  & 15     & -2.4 & 17  &  21        &  63 \\
    93  & 50  & 25     & -2.6 & 1.9 &  2.6       &  30 \\
    145 & 50  & 25     & -3.0 & 2.1 &  3.3       &  17 \\
    225 & 70  & 35     & -3.0 & 4.2 &  6.3       &  11\\
    280 & 100 & 40     & -3.0 & 10  &  16        &  9 \\
    \hline
    \hline
\end{tabular}
\end{table*}

\subsection{Sky area and expected noise levels} \label{sec:noise}

As described in SO19, the Simons Observatory plans to conduct two surveys: a
large survey covering about 40\% of the sky, conducted by the LAT, and a smaller
survey of the cleanest \(\sim 10-20 \%\) of the sky using the SATs
\citep{stevens/etal:2018}.

The noise model used in this study is described in detail in SO19, and we
summarize it here. Two noise levels are considered: the `baseline' design
assumes a raw sensitivity based on the achieved performance of previous ground
based experiments such as {\sc ABS} \citep{kusaks/etal:2018}, {\sc BICEP}
\citep{bk:2016}, and {\sc QUIET} \citep{quiet:2011,quiet:2012}, and a `goal'
design which will require more abitious detector development. These raw
sensitivities are multiplied by an observing efficiency of 20\%, accounting
for all data cuts, observing downtime and instrument calibration, based on the
efficiency achieved during observations at the same site by the ACT experiment.

\begin{figure*}[t]
  \centering
  \includegraphics[]{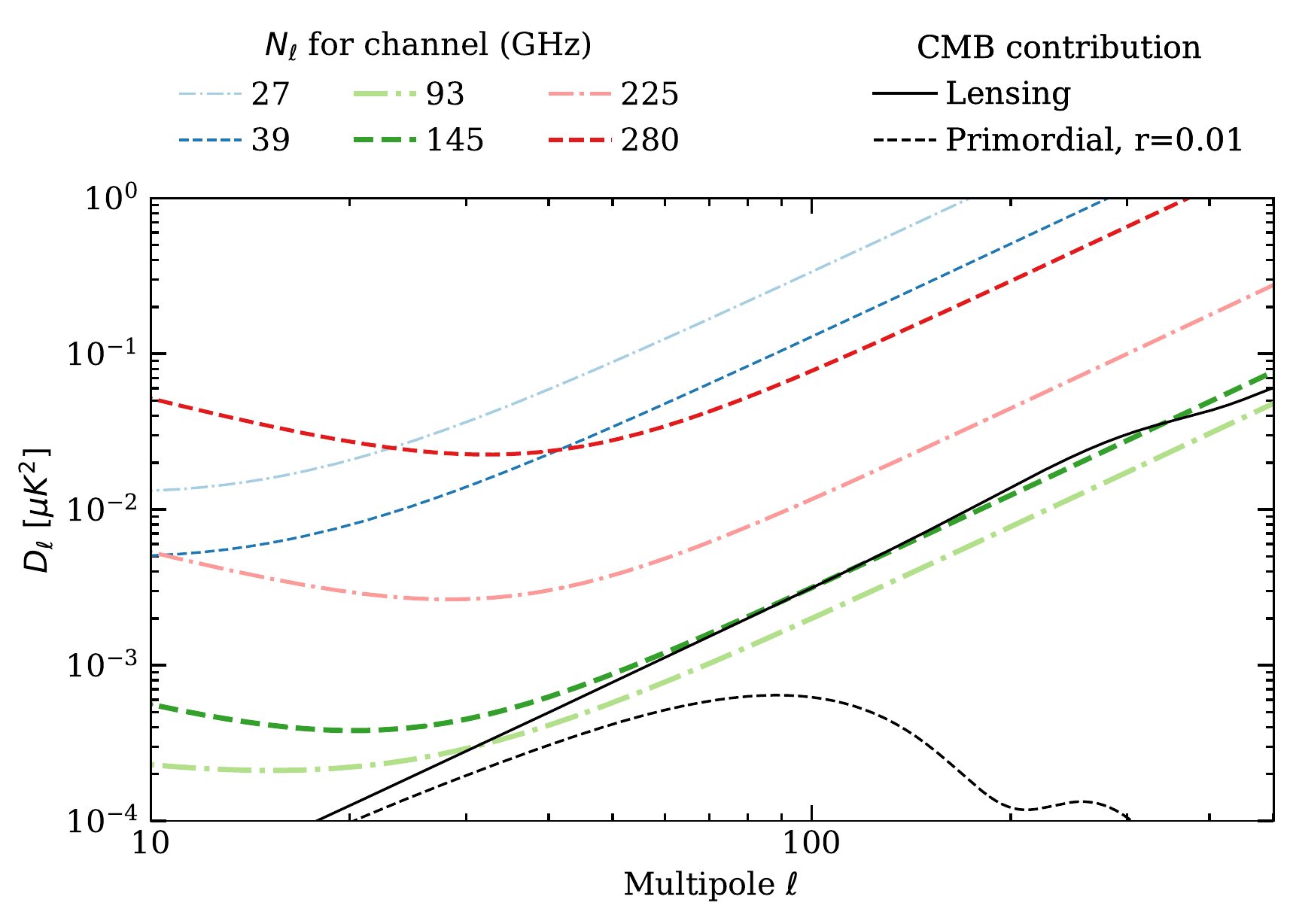}
  \caption{Expected noise curves for SAT polarized observations, from
  	\citet{so_forecasts:2018}, for baseline sensitivity and optimistic
  	$\ell_{\rm knee}$ parameter. The lensing power spectrum is also shown for
  	comparison.}
    \label{fig:bb_noise_curves}
\end{figure*}

The $1/f$ noise induced in the insrument by atmospheric loading and instrument
systematics is parameterized as an additional term in the noise power spectrum
that increases at large scales:
\begin{equation}
  \label{eq:instrument_noise}
  N_{\ell} = N_{{\rm white}}\left[1 + \left(\frac{\ell}{\ell_{\rm knee}}\right)
  ^{\alpha_{\rm knee}}\right],
\end{equation}
where $\alpha_{\rm knee}$ and $\ell_{{\rm knee}}$ are the knee index and
multipole, respectively. The range of parameters we consider here, as in SO19,
are summarized in Table \ref{tab:somatrix}. In Figure \ref{fig:bb_noise_curves}
we show the individual frequency noise curves, from SO19, for the optimistic
knee multipole, and goal sensitivity, compared to the lensing $B$ mode spectrum.

Due to the large field of view of the SAT, the survey design has non-uniform
depth. Therefore, in modeling the noise properties it is important to weight
noise realizations by the relative hit density. We follow the same procedure as
in SO19 to generate noise realizations with this non-uniform hit density, and
non-white noise.

In SO19, the effect of the resolution of the lowest two frequency channels
on constraints on $r$ was studied. It was found that the impact of the
resolution of these channels did not affect the achieved sensitivity
dramatically. In order to simplify the later analysis, we therefore make the
assumption that all frequencies have the same $30^\prime$ FWHM resolution.

We generate 200 Monte Carlo realizations of the noise and CMB, and use a common
foreground realization for this suite of simulations.

\begin{figure*}[t]
  \centering
  \includegraphics[]{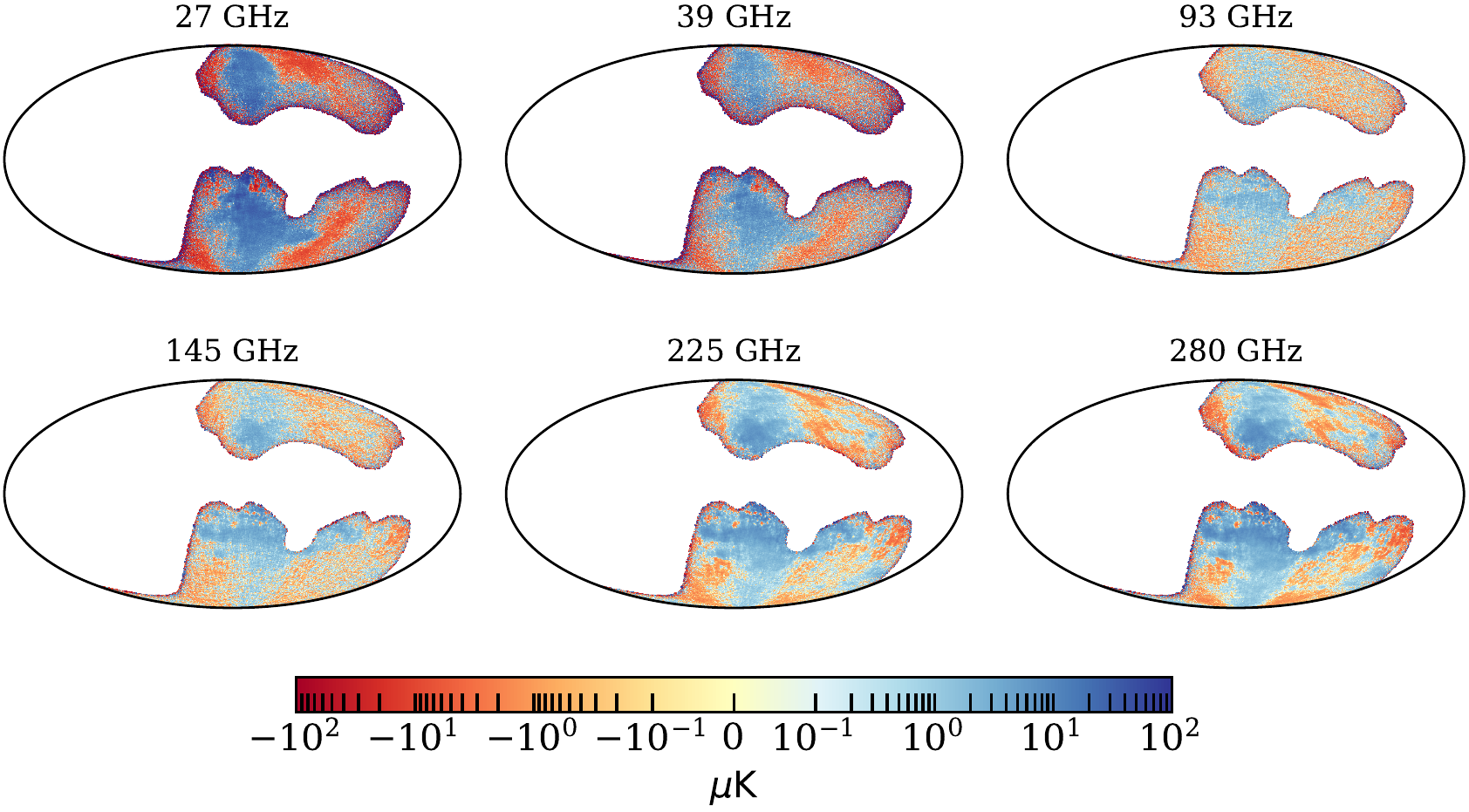}
  \caption{Maps in Galactic coordinates showing simulations of the $Q$ Stokes
  	parameter for {\sc simset1} with the baseline sensitivity and optmistic
  	$\ell_{{\rm knee}}$ configuration of SO. Note that the colorscale is a
  	combination of a linear scale between $(-0.1, 0.1)~{\rm \mu K}$ and a log
  	scale outside this range, to display structure over a large dynamic range.
    In the 27 GHz channel, the large scale structure of synchrotron emission
    is visible, with the North Galactic Spur clearly visible above the Galactic plane. At the CMB frequencies 93 GHz and 145 GHz, the CMB $E$ mode
    polarization becomes visible at high Galactic latitudes. At the highest
    frequency channels the morphology of dust
    emission becomes dominant.}
  \label{fig:example_so_obs}
\end{figure*}

\section{Component Separation and parameter estimation}
\label{sec:component_separation}

In this section we describe the map-space component separation algorithm, and
the subsequent estimation of power spectra and parameters from the cleaned maps.

\subsection{Component separation}
Our component separation method follows the {\tt BFoRe} method described in
\citet{alonso17_simul_forec_primor_b_searc}. We take the simulated sky maps,
${\bf d}$, and model them as a linear combination of
components with spatially varying SEDs, and noise:
\begin{equation}
  {\bf d}(\hat n) = {\sf F}({\bm \theta}(\hat n)) \cdot {\bf T}(\hat n) +
  {\bf n}(\hat n),
\end{equation}
where ${\sf F}({\bm \theta}(\hat n))$ is the mixing matrix containing different
component SEDs, with parameters ${\bm \theta}(\hat n)= \{{\bm \beta}_{d}(\hat n),
{\bf T}_{d}(\hat n), {\bm \beta}_{s}(\hat n)\}$, ${\bf T}(\hat n)$ is a vector of
component templates at a specific reference frequency, and ${\bf n}$ is a
noise term. Comparing to \ref{eq:sim_eq} we see that this is essentially the
`correct' model, modulo the permitted degree of spatial variation of the spectral
parameters ${\bm \theta}(\hat n)$, and beam-convolution.

Under the assumption of Gaussian noise we can write down the likelihood for the data:
\begin{equation}
  - 2 \ln [\mathcal{L(\mathbf{T, {\bm \theta} | \mathbf{d}})}] \propto
  ({\bf d} - {\sf F} \cdot {\bf T})^T {\sf N}^{-1} ({\bf d} - {\sf F} \cdot
  {\bf T})
  \label{eq:lkl}
\end{equation}
where \({\sf N}^{{-1}}\) is the data covariance. We simplify the analysis by
assuming that the noise is diagonal in pixels and frequencies, however we
emphasize that the true noise in non-white, making this approach sub-optimal.
This assumption allows us to write the covariance as a diagonal matrix:
\begin{equation}
  ({\sf N}^{{-1}})_{(i, j, \nu), (i^{\prime}, j^{\prime}, \nu^{\prime})} =
  \sigma_P^2~\delta_{(i, j, \nu), (i^{\prime}, j^{\prime}, \nu^{\prime})}.
\end{equation}
Under this assumption, the likelihood can be separated into a product over
$N_{\rm spec}$ large pixels in which the spectral parameters are allowed to
vary. In principle, these large pixels are not tied to any particular
pixelization scheme, but can be implemented as any arbitrary shape intended to
follow the true spatial variation of spectral parameters. The complexity of
the foreground model can be increased by allowing more parameters to vary in the
fit, or by increasing the number of independent patches.

Unless otherwise stated, during the rest of this paper we maximize the likelihood
in equation \ref{eq:lkl} by varying only the dust and synchrotron spectral
indices, ${\bm \beta}_d(\hat n)$ and ${\bm \beta}_s(\hat n)$, keeping the
temperature fixed at $20~{\rm K}$. We carry out this maximization in each large
pixel, and for each Monte Carlo realization of CMB and noise. This is equivalent
to marginalizing over the spectral indices for a single realization.

\subsection{$B$ mode power spectrum estimation}
\label{sec:estimation_of_r}

In order to constrain the tensor-to-scalar ratio we first calculate the power
spectrum from the cleaned CMB ${\bf Q}(\hat n)$ and ${\bf U}(\hat n)$ maps. The transformation from
${\bf P} = ({\bf Q}(\hat n), {\bf U}(\hat n))$ to $\mathbf{P}^\prime_{\ell m} = ({\bf E}_{\ell m}, {\bf B}_{\ell m})$ is inherently non local as it requires
the calculation of spherical harmonic transforms:
\begin{equation} \label{eq:pseudospectrum}
  {\bf P}^{\prime}_{\ell m} = \int_{{4 \pi}} d\Omega \ {\sf Y} \cdot {\bf P},
\end{equation}
where ${\sf Y}$ is a $2 \times 2$ matrix, with each element being a specific
combination of spin-weighted spherical harmonics
\citep[e.g.,][]{grain/etal:2009}.

Ground based instruments observe only part of the sky, and so can only access
the true sky, multiplied by some window function, ${\bf w}(\hat n)$:
$\tilde{{\bf P}}(\hat n) \equiv ({\bf w}(\hat n){\bf Q}(\hat n),
{\bf w}(\hat n){\bf U}(\hat n)$. A naive calculation using the standard
pseudo-power spectrum technique \citep{wandelt/etal:2001} will mix $E$ and $B$
modes, and if not accounted for increases the variance of the estimated
$B$ modes, limiting the achievable constraints on the tensor-to-scalar ratio
\citep{challinor/chon:2005, smith:2006}.

To correct for this effect we use a `pure' estimator of the power spectrum
\citep{smith:2006, smith/zaldarriaga:2007, grain/etal:2009}, which is equivalent
to first calculating the naive pseudo-spectrum of the maps over some mask,
${\bf w}(\hat n)$, and then calculating and removing the leaked $E$ modes
\citep{grain/etal:2009}. This method assumes that the applied mask satisfies
Dirichlet and Neumann boundary conditions. Therefore, we apply an additional
tapering to the inverse variance map that would usually be used in the
calculation of the power spectrum. In this work we use the publicly available
{\tt NaMaster}\footnote{\url{https://github.com/LSSTDESC/namaster}} code.
For details of the implementation see the {\tt NaMaster} documentation and
\citet{alonso/etal:2019}, and for further theoretical background see
\citet{smith:2006, smith/zaldarriaga:2007, grain/etal:2009}.

Due to the limited sky coverage, there is insufficient information to invert the
mode coupling matrix for all multipoles. Instead we coarse-grain the matrix by
defining some binning scheme. In this work we use the binning operator:
\begin{equation}
\label{eq:binning_scheme}
  {\sf W}_{\ell_b \ell} = \frac{1}{\Delta \ell} \Theta(\ell - \ell_{b})
  \Theta(\ell - \ell_{b} + \Delta \ell),
\end{equation}
where $\Theta$ is the Heaviside function, $\ell_{b}$ denotes the bandpower, and
$\Delta \ell$ is the width of each bin, which we choose to be $\Delta \ell=5$.
Then estimates of the binned power spectrum are:
\begin{equation}
C_{\ell_{b}}^{XX} = \sum_{\ell}{\sf W}_{\ell_{b} \ell}\sum_{m}\frac{|\tilde{a}^{XX}_{\ell m}|^{2}}{2\ell + 1 }.
\end{equation}

From the cleaned Monte Carlo simulations we calculate the expected power
spectrum, and its covariance. In order to avoid noise bias, we cross-correlate
only cleaned maps with different noise realizations.

\subsection{Cosmological parameter likelihood}
The full posterior for the individual bandpowers is non-Gaussian. However, for high enough
multipoles the central limit theorem justifies a Gaussian approximation \cite{hamimeche/lewis:2008;}.
The combination of the compact observing region and atmosphere-induced systematics, limit
constraints on large scale modes.  Therefore, we consider only $\ell > 30$, for which the
Gaussian approximation is valid:
\begin{widetext}
  \begin{equation}
    \begin{aligned}
      - 2 \ln[ \mathcal{L}(r, A_L)] &= (\hat C^{BB}_{\ell_b}-C^{BB}_{\ell_b}(r, A_L))^T {\rm Cov}
      (\hat C^{BB}_{\ell_b}, C^{BB}_{\ell_b})^{-1}(C^{BB}_{\ell_b}-C^{BB}_{\ell_b}(r, A_L)), \\
      C^{BB}_{\rm \ell_b} &= r C_{\ell_b}^{BB, {\rm prim}}(r=1) +
        A_{L}C_{\ell_{b}}^{BB, {\rm Lens}}(A_{L}=1),
    \end{aligned}
    \label{eq:bb_lkl}
  \end{equation}
\end{widetext}
where the measured power spectrum is indicated by the hat, and the model,
$C^{BB}_{\ell_b}$, is the sum of a primordial term and a lensing term,
$C_{\ell_{b}}^{BB, {\rm Lens}}(A_{L}=1)$ is a template for the lensing
contribution, and $C_{\ell_{b}}^{BB, {\rm prim}}(r=1)$ is a template for the
primordial contribution, both of which are calculated using the {\tt CLASS} code
with Planck 2018 cosmological parameters \citep{planck_cosmo:2018} and $r=1$.
Equations~\ref{eq:bb_lkl} are given at the bandpowers $\ell_b$, having
accounted for the effects of bandpower averaging and inversion of the mode
coupling matrix \citep[e.g.,][]{grain/etal:2009, alonso/etal:2019}.

We sample this likelihood for these two parameters using the
{\tt emcee} package, and summarize the posterior on $r$ by marginalizing over
the lensing amplitude and calculating the median and standard deviation of the
posterior distribution.

\section{Results}
\label{sec:results}

In this section we present the results of applying the foreground separation
algorithm described in Section \ref{sec:component_separation} to the simulations
described in Section \ref{sec:simulations}.

We first present the results for {\sc simset1}, which are the fiducial sky
simulations analyzed in SO19. As noted, the foreground-cleaning algorithm
presented in this paper is similar to the {\tt BFoRe} method used in
\cite{so_forecasts:2018} and \cite{alonso17_simul_forec_primor_b_searc}, but uses
an independent pipeline. Therefore, we first validate our analysis, and then
provide an extension of the cases studied in previous works.

\subsection{Fiducial sky simulations}

\begin{figure}[t]
  \includegraphics[]{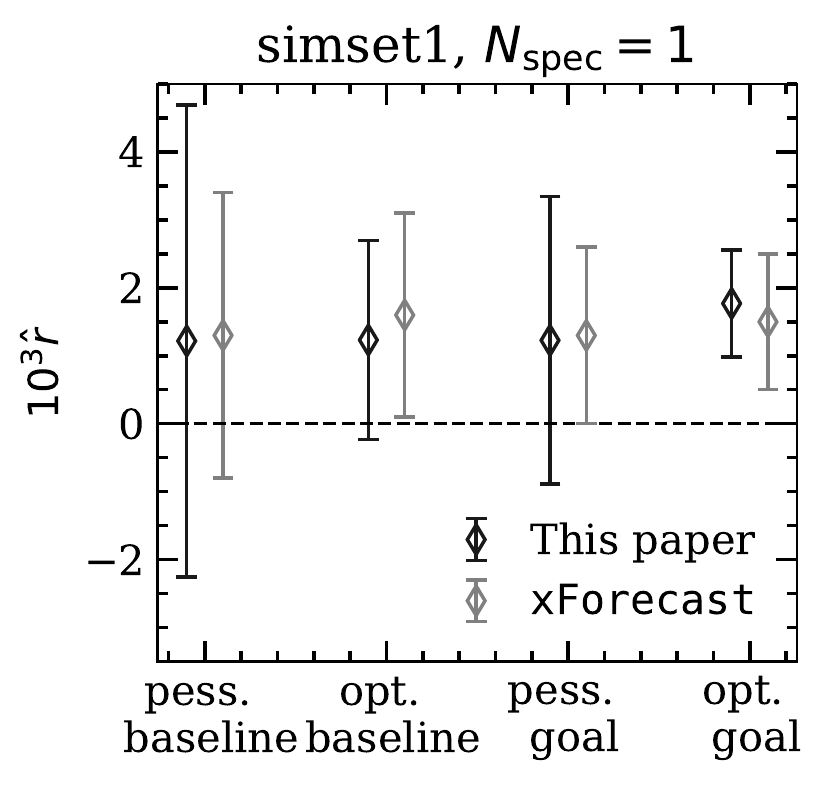}
  \caption{Forecasted constraints on $r$, for a model with $r=0$, from simulated
  	observations of {\sc simset1} for the four Simons Observatory noise levels.
  	The results from the 200 Monte Carlo simulations in this work, assuming no
  	spatial variation of the spectral indices (black), are compared to results
  	from the {\tt xForecast} method presented in SO19 (grey).}
  \label{fig:simset1_vary_instrument}
\end{figure}

We show the resulting bias and uncertainty on $r$ for the four Simons Observatory
noise configurations in Figure \ref{fig:simset1_vary_instrument}, compared to the
results of the {\tt xForecast} \citep{errard/stompor:2018} framework presented in
SO19.  We compare to {\tt xForecast}, an alternative parametric method that uses
the same simulations, because the reported BFoRe forecasts in SO19 include an
additional marginalization over residual foregrounds while estimating the
tensor-to-scalar ratio. This comparison is therefore done for the specific
{\tt xForecast} case where no additional foreground power is marginalized over.
Figure \ref{fig:simset1_vary_instrument} shows that our method produces
consistent forecasts with those in SO19. The nominal Simons Observatory design is
biased by 1$\sigma$, and the most sensitive design is biased by 2$\sigma$. The
errors from our method and {\tt xForecast} are not identical; we have checked
(private communication) that our forecasted errors agree with the {\tt BFoRe}
code in the case of no-marginalization.


In the rest of this section we focus on the case of optimistic $\ell_{\rm knee}$
and baseline sensitivity. We demonstrate that the source of the bias in $r$ is
due to foreground contamination, primarily from residual dust. In SO19 the bias
was largely removed at the power spectrum level by marginalizing over a template
power spectrum for the foreground contamination. Relying on this method for an
unbiased detection of primordial gravitational waves could be problematic as the
shape of the residual foreground power spectrum is not known {\it a priori}.

In the rest of this section we establish that neglecting spatial variation of the
indices is the root cause of the bias, and that it may be removed with additional
masking of the higher-foreground regions, or by introducing only a few more
parameters describing the spatial variation of the dust spectral parameters.

\subsection{Establishing source of bias}

\begin{figure*}[t]
	\includegraphics[width=0.49\textwidth]{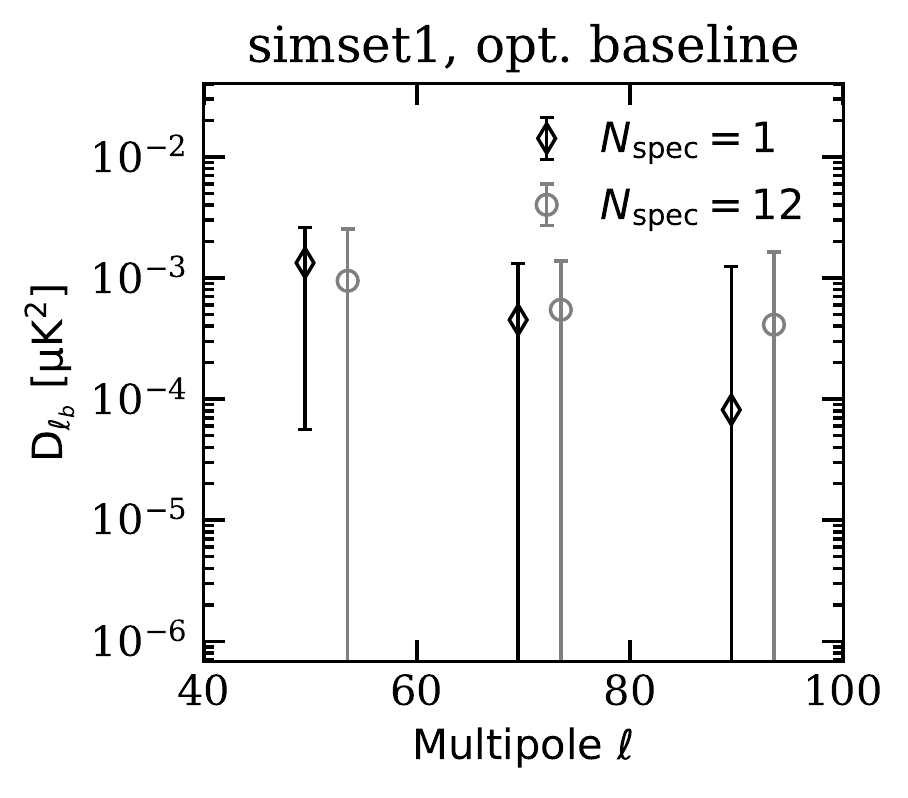}
	\includegraphics[width=0.49\textwidth]{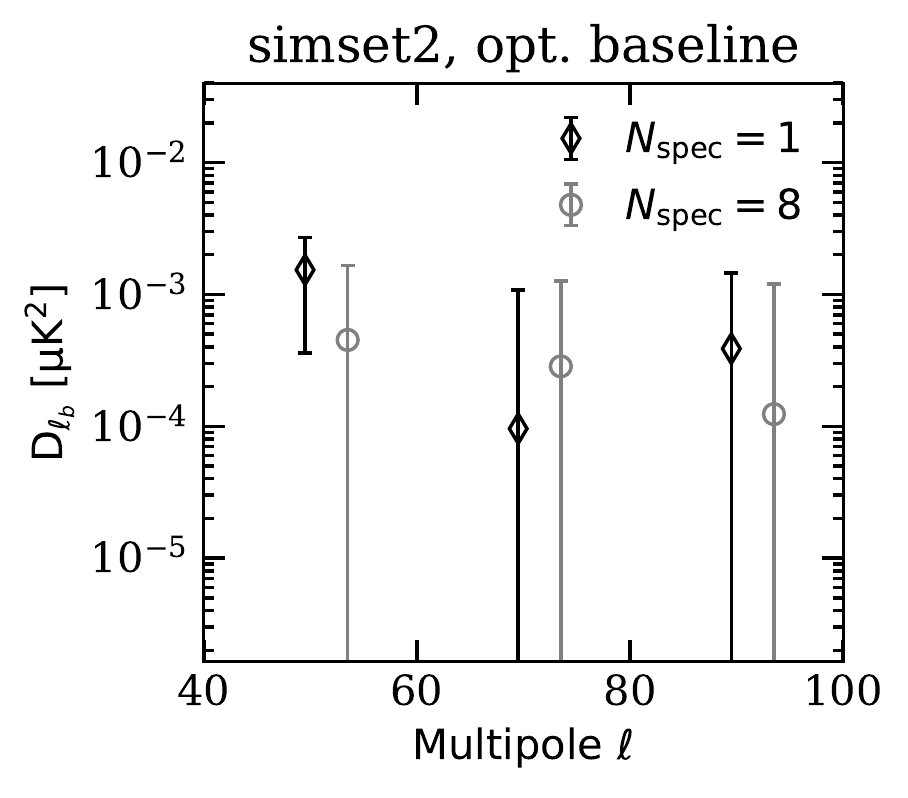}
	\caption{The cross-correlation of cleaned CMB maps with Galactic templates can
		be used to test for residual foregrounds. This shows the cross-spectrum when
		cleaning with a spatially constant dust SED and synchrotron spectral index
		(diamonds) or accounting for spatial variation (circles). We see a non-zero
		signal at the largest scales when using the spatially constant models, which
		correlates with a bias in $r$. (Left) This uses {\sc simset1}, and correlates
		with a template of Galactic dust. (Right) This uses {\sc simset2} and
		correlates with a synchrotron template.}
	\label{fig:xcorr}
\end{figure*}

\subsubsection{Masking the Galactic plane}

To establish the root of a non-zero value of $r$, the result must be
robust against different splits of the data. In this section we establish that
increasing the masking of the Galactic plane when estimating the power spectrum
from cleaned CMB maps reduces the spurious detection of $r$, and we show that
sufficient masking can remove all of the bias in $r$, at the expense of a small
increase in the uncertainty.

\begin{figure}[t]
  \includegraphics[]{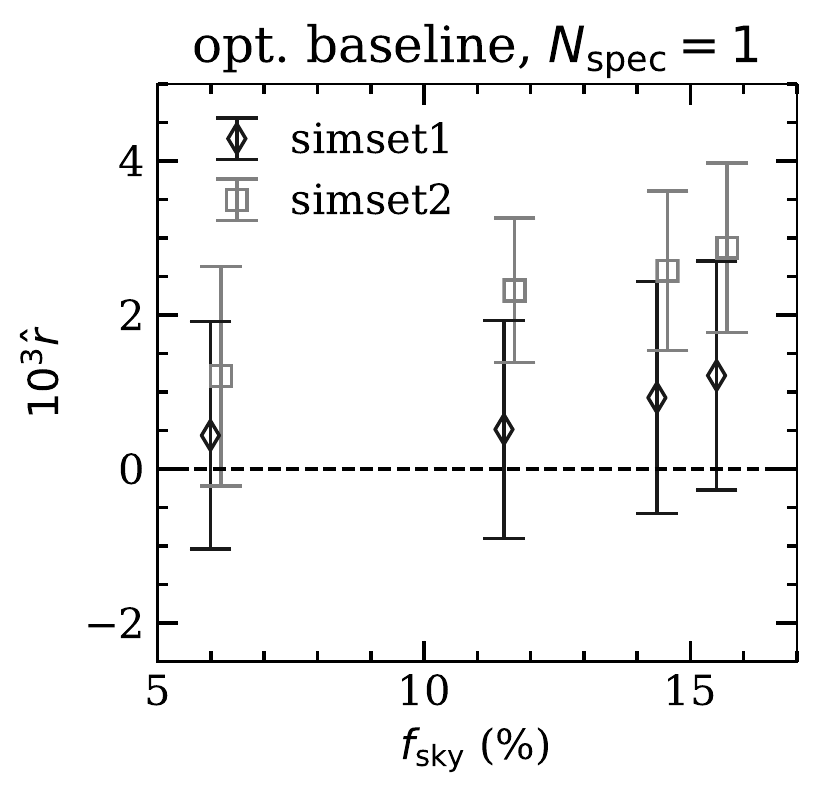}
  \caption{Forecasted constraints on $r$ as a function of sky area, from
  	simulated observations of {\sc simset1} (diamonds) and {\sc simset2}
  	(squares), when foregrounds are removed assuming no spatial variation of
  	spectral indices. Each point represents a different level of Galactic masking
  	when computing the power spectrum of the cleaned CMB maps. The bias is
  	reduced when the brightest sky regions are masked.}
    \label{fig:simset1_so11_galmas}
\end{figure}

We use a set of Galactic masks based on thresholds of the Planck intensity
maps at 353 GHz, downloaded from the Planck Legacy Archive
\footnote{
	\href{http://pla.esac.esa.int/pla/aio/product-action?MAP.MAP_ID=HFI_Mask_GalPlane-apo0_2048_R2.00.fits}{Filename:
		{\tt HFI\_Mask\_GalPlane-apo0\_2048\_R2.00.fits}}}.
The masks range from leaving 20 \% to 80 \% of the sky unmasked. We combine each
Planck mask with the SO hits map from SO19, resulting in four maps with
effective sky area from $\sim 6$\% to $\sim 16 \%$, calculated using the
appropriate expression for noise-dominated maps \citep{so_forecasts:2018}:
\begin{equation}
 f_{\rm sky} = \frac{\langle w^2 \rangle^2}{\langle w^4 \rangle},
\end{equation}
where $w(\hat n)$ is the combined hit map, and Galactic mask. For each mask we
repeat the power spectrum estimation and cosmological parameter fitting to
estimate $r$, and present the results in Figure \ref{fig:simset1_so11_galmas}. As
the masks become more aggressive, the bias in $r$ is reduced substantially for
both {\sc simset1} and {\sc simset2}. Since we expect foreground residuals to be
localized to the Galactic plane, a decreasing bias with increased plane masking
indicates the Galactic nature of the bias.

\subsubsection{Cross-correlating with foreground templates}

If the cleaned CMB maps are contaminated by foreground residuals, we may expect
there to be a significant correlation with templates of the individual foreground
components. In order to perform this test using only the observed data, without
relying on external observations, we form templates of high and low frequency
foregrounds in the following way.

A synchrotron template is constructed by differencing the 27 GHz and 40 GHz maps,
and a dust template is constructed by differencing the 270 GHz and 220 GHz maps.
In thermodynamic units the CMB contribution will cancel, and we will be left with
a map proportional to the synchrotron and dust populations, respectively. The
foreground templates are then correlated with the cleaned CMB maps. We may expect
the noise in the constructed templates and cleaned CMB maps to be correlated,
leading to biases in the recovered power spectra. Therefore, we only
correlate templates and CMB maps from different half-mission splits of the data.
Recent work  \citep[e.g.,][]{clark/etal:2015, clark:2018} has shown that linear
Galactic neutral hydrogen features are correlated with polarized Galactic
foregrounds. Maps of HI emission could provide an independent test of polarized
dust contamination in cleaned CMB maps.

In Figure \ref{fig:xcorr} we show the cross spectra of {\tt simset1} and
{\tt simset2} with dust and synchrotron templates, respectively, using the
largest 16\% sky region and cleaning with a single synchrotron and dust index. We
find that the residual foregrounds, which showed up as a bias in $r$ in Figure
\ref{fig:simset1_vary_instrument}, can be detected in the cross-spectra.

From these exercises we find that: i) a single spectral parameter is insufficient
to describe the spectral energy dependence of foregrounds in the region observed
by the Simons Observatory, and results in a bias $\ge 1.5 \times 10^{-3}$ for the
most optimistic case of foreground complexity, as was found in SO19.
ii) depending on the complexity of the true foregrounds, the bias can be
primarily due to  dust or to synchrotron contamination, iii) the bias can be
mostly removed by using only $\sim$ 6 \% of the sky. These conclusions are true
in so far as these simulations are representative of the true sky, which may
in fact be more complicated.

\subsection{Spatially varying spectral indices}

In this sub-section we demonstrate that the residual foreground contamination
occurs because the fitting process does not account for the spatial variation of
the spectral indices, and we demonstrate that such variation may be sufficiently
described by only a few additional parameters if they could be chosen to
accurately capture the spatial variation.

We adapt the map-space cleaning algorithm to account for spatial variation of the
foregrounds by defining a set of $N_{\rm spec}$ patches in which to fit spectral
indices. In theory $1 \leq N_{\rm spec} \leq N_{\rm pix}$, where $N_{\rm spec} =
1$ corresponds to fitting a single spectral index over the whole sky, and
$N_{\rm spec}=N_{\rm pix}$ fits a spectral index in every pixel at the resolution
of the input maps. The limit of large $N_{\rm spec}$ would account for as much
spatial variation in the foreground SEDs as possible, however requires increasing
the number of fitted parameters, and therefore post-separation noise. It would
therefore be ideal to keep $N_{\rm spec}$ as low as possible, whilst still
achieving an unbiased estimation of $r$. These simulations were generated at
$N_{\rm pix}$ resolution, therefore, there would be no advantage to choosing
$N_{\rm spec}>N_{\rm pix}$. Note that, for observations of the real sky, the
creation of pixelized maps requires spatial averaging both along the line of
sight, and transverse to it. Such averaging could be better fitted by taking
$N_{\rm spec}>N_{\rm pix}$, as the extra parameters could absorb some of the
resulting SED curvature \citep{chluba/etal:2017, krachmalnicoff/etal:2018}.
Recent work \citep[such as the moment-based expansion of ][]{chluba/etal:2017}
has suggested introducing additional parameters in the SED to capture the
effects of spectral variations at smaller scales. This would avoid the
complication of deciding how to subdivide the sky. Comparison to this
approach would be interesting, and is left for future work.

How to distribute patches on the sky is an important consideration, as it
involves a choice of how to model the spatial dependence of foreground
parameters. This is unknown {\it a priori}, so we explore an initial exploration
of possible approaches.

\begin{figure*}[t]
	\centering
\includegraphics[]{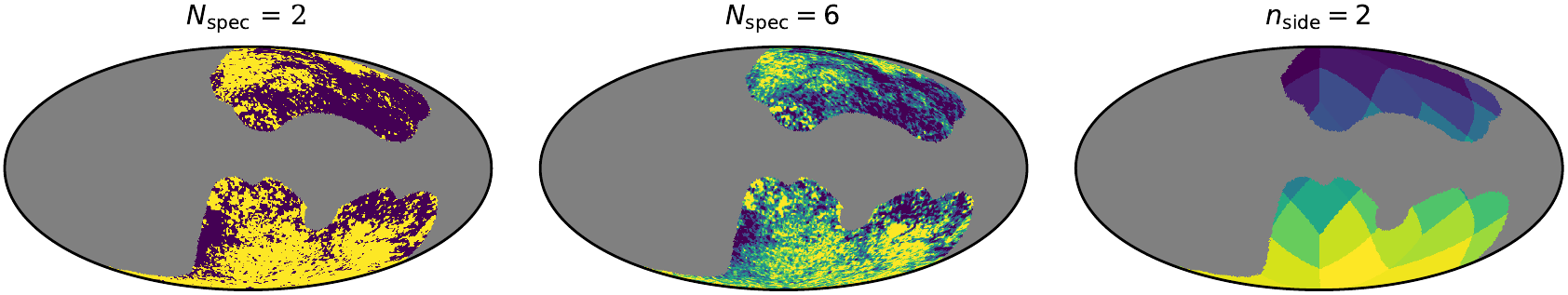}
 \caption{(Left and middle) The fitting regions used to fit different dust SEDs,
 	as defined by the algorithm described in  \ref{sec:betad_grid}, for
 	$N_{\rm spec}=2, 6$. (Right) The fitting regions using a {\tt Healpix}
 	$n_{\rm side}=2$ grid.}
 \label{fig:betad_fitting}
\end{figure*}

\subsubsection{Fitting spectral indices on a {\tt HEALPix} grid}
\label{sec:healpix_grid}

\begin{figure}[t]
	\includegraphics[]{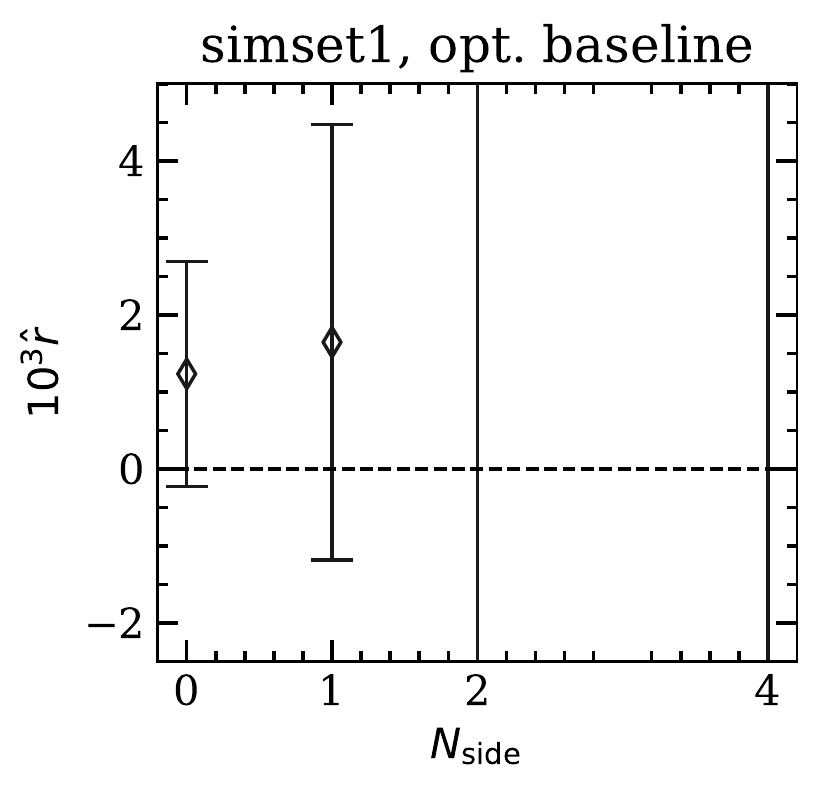}
	\caption{Forecasted constraints on $r$ when fitting spectral indices defined
  	by {\tt HEALPix} grids at increasing resolution. Since SO observes a
  	fraction of the sky, the number of coarse pixels is given by
  	$12 f_{\rm sky} n_{\rm side}^2$. As $n_{\rm side}$ increases we find the
  	projected uncertainty on $r$ increases significantly: this is not a good
  	choice for dividing up the sky area.}
    \label{fig:nominal_hpix}
\end{figure}

Some previous studies have used patches corresponding to a coarse {\tt HEALPix}
grid for convenience \citep[e.g.,][]{alonso17_simul_forec_primor_b_searc},
allowing an independent spectral parameter to be fit in each coarse pixel,
as shown in the right panel of Figure~\ref{fig:betad_fitting}.
We define a set of fitting models with spectral parameters varying on grids at
$n_{\rm side}=1, 2, 4$ (with $n_{\rm pix}=48, 192, 3072$), and perform the
foreground cleaning. We show our resulting constraints on $r$
in Figure~\ref{fig:nominal_hpix}. We find that the errors are significantly
inflated as $n_{\rm side}$ increases, but determine that this is partly due to
many of the larger {\tt Healpix} pixels having too few observed pixels, and also
due to having so many additional parameters.

\subsubsection{With prior information on spatial variation of indices}
\label{sec:betad_grid}

\begin{figure}
\begin{algorithm}[H]
		\caption{This algorithm is used to produce a set of fitting regions for
			an input spectral  index map, $\beta$, and integer $N_{\rm spec}$.}
			\label{alg:1}
	\begin{algorithmic}[1]
		\STATE Resample $\beta$ at $n_{\rm side}=256$ to obtain the map
		$\beta_{256}(i)$, where $i$ is the pixel index .
		\STATE Restrict to the subset of pixels observed by SO,
		$\beta_{256}^{SO}$.
		\STATE Create histogram of $\beta_{256}^{SO}$ with $N_{\rm spec}$
		equally  sized bins, with edges $\beta_{j}$, where
		$j = 0, ..., N_{\rm spec}$.
		\STATE Create an empty $n_{\rm side}=256$ map, $b(i)$.
		\STATE For each pixel in $b$ assign $b(i) = j$, where $j$ is the number
		of the bin into which $\beta_{256}^{SO}(i)$ falls.
		\STATE Return $b(i)$.
	\end{algorithmic}
\end{algorithm}
\end{figure}

With prior information about the spatial variation of the spectral parameters
we investigate how many additional parameters would be needed to mitigate the
observed biases in $r$ due to foreground residuals. Here we use the dust
spectral index map, ${\bm \beta}^{\rm dust}_{d1}(\hat n)$, used in {\tt PySM}
model `d1', to create a template that defines the regions in which we fit
spectral parameters. The method used to define $N_{\rm spec}$ regions from
${\bm \beta}_{d1}^{\rm dust}(\hat n)$ is described in detail in
Algorithm~\ref{alg:1}.

We repeat Algorithm~\ref{alg:1} for $N_{\rm spec} \in 1, ..., 12$, to produce
twelve different template maps, with increasing spatial resolution. Figure
\ref{fig:betad_fitting} shows the regions for $N_{\rm spec} = 2, 6$.

\begin{figure*}[t]
	\includegraphics[width=0.49\textwidth]{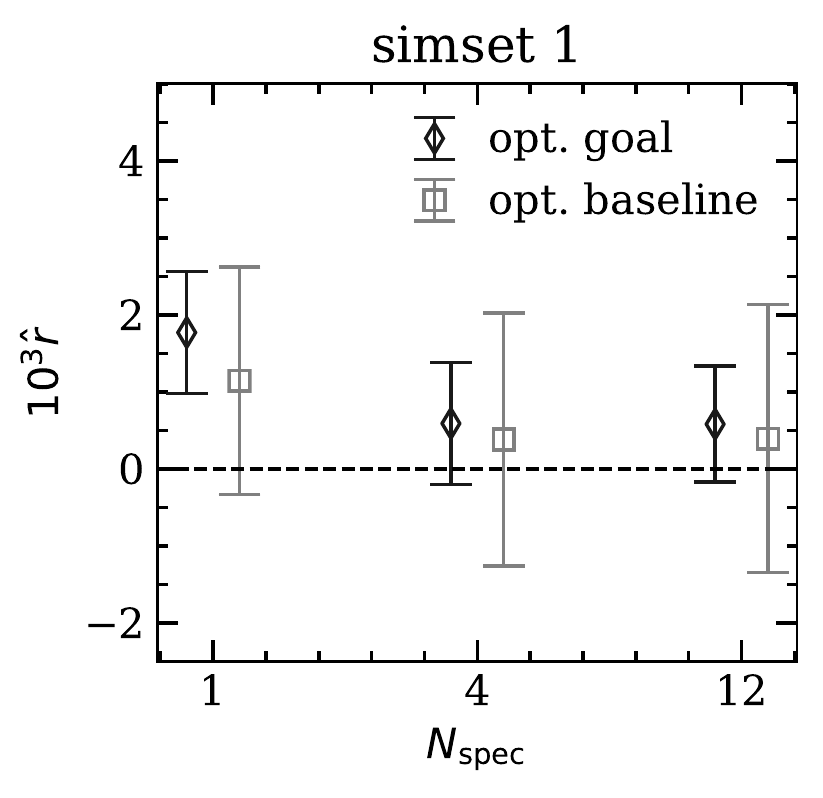}
	\includegraphics[width=0.49\textwidth]{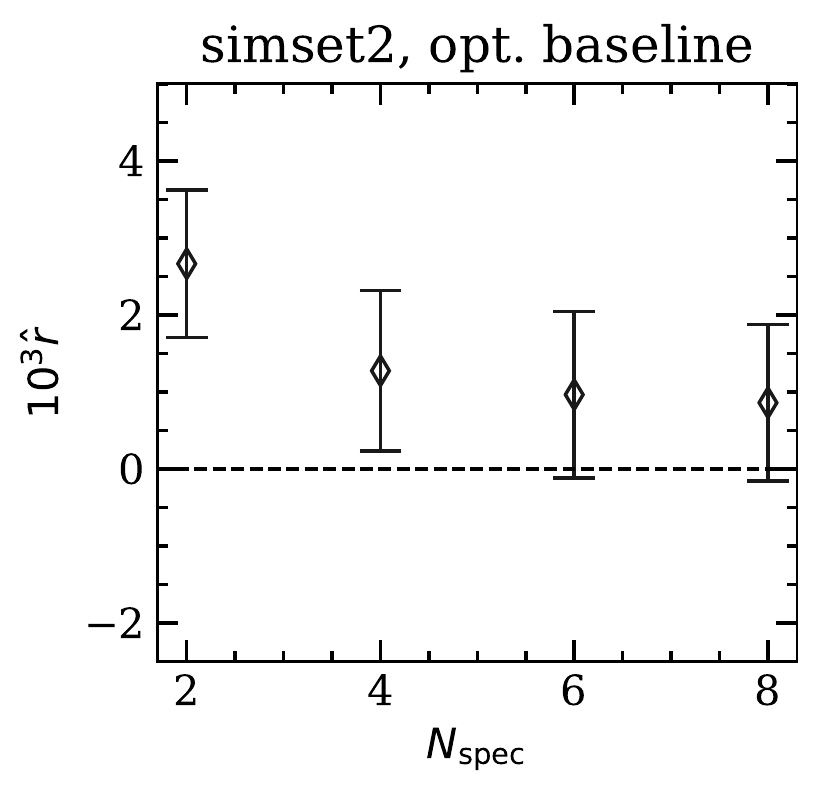}
	\caption{(Left) Forecasted constraints on $r$ when fitting models with
		increasing independent regions defined from the $\beta^{\rm dust}_{d1}$
		template, for the optimistic $\ell_{\rm knee}$ goal and baseline
		sensitivities. The bias is removed with a modest increase in parameters,
		if these regions are known a-priori. Right: using {\sc simset2} and
		fitting models with regions defined from the $\beta^{\rm sync}_{s2}$
		template, a similar effect is seen}
      \label{fig:index}
\end{figure*}

We run our foreground cleaning algorithm on the {\sc simset1} sky simulations
for these spectral index pixelizations described above. The resulting
constraints we find on $r$ are shown in Figure \ref{fig:index} for
$N_{\rm spec}=1, 4, 12$. We find that increasing $N_{\rm spec}$ results in a
reduced bias on $r$, as we are able to account for spatial variation of
$\beta_d$, with negligible inflation of the uncertainty. No improvement of the
bias is achieved past a value of $N_{\rm spec} \approx 4$. Since the fitting code
is designed to fit all parameters in the same patches, this remaining bias is due
to the mismatch between the dust-focused fitting regions and synchrotron spectral
index variation. Modifying the fitting code to allow different physical
parameters to be fitted with different constraints is a non-trivial extension,
and is left to future work.

In Figure \ref{fig:xcorr} we show the cross-correlation of the cleaned CMB maps
with the dust foreground templates, as a function of number of fitting pixels
$N_{\rm spec}$. Increasing the number of fitting regions results in a better
cleaning  of the foregrounds, and consequently we find no signature of residual
contamination by $N_{\rm spec}\sim4$.

We then repeat this study for {\sc simset2}, which has a synchrotron spectral
index with a greater spatial variance, $\beta^{\rm sync}_{s2}$. Ignoring spatial
variation of the foreground SED in this case results in an even larger bias on
$r$ than in the case of {\sc simset1}. Here we use $\beta^{\rm sync}_{s2}$ to
define the sky regions used for fitting the indices, and $N_{\rm spec}=2, ..., 8$
as the input for Algorithm~\ref{alg:1}. Our resulting constraints on $r$ are
shown in Figure~\ref{fig:index}. Again, there is a reduction in the bias on $r$
as we increase $N_{\rm spec}$, and the improvement saturates at
$N_{\rm spec}\approx 6$. For $N_{\rm spec} \gtrsim 6$ we see no improvement as
the bias is now dominated by residuals due to dust mis-modeling. Indeed the
remaining bias for $N_{\rm spec} \gtrsim 6$ is at the
$\sim 1 \sigma$ level found when fitting {\sc simset1} with a single spectral
index.

In Figure~\ref{fig:xcorr} we show the correlation of the cleaned CMB maps with a
synchrotron template formed from the two lowest frequency channels. Similar to
{\sc simset1} we find that increasing $N_{\rm spec}$ removes the hints of
foreground bias at low multipoles seen when ignoring spatial variation of
spectral indices.

It is promising that with only 4-6 additional parameters describing the spatial
variation of the synchrotron and dust indices, that a 1-3$\sigma$ bias in $r$
might be mitigated without significant masking. In practice, though, we will not
have the perfect information about their spatial variation. There may also be
additional variation not captured in the existing simulations. Instead, we would
need to derive the fitting regions directly from the data.

Towards this, we determined that the $\chi^2$ of the model compared to the data,
with the intent to add spatial resolution to the fitting model in areas of poor
$\chi^2$, was not sufficiently sensitive to discriminate between areas of good
and poor fit. A judicious choice of these fitting regions, based on existing and
upcoming observations, will be a natural direction for future study.

\section{Conclusions}
\label{sec:conclusions}

By accounting for Galactic foreground models based on current
data, we have assessed how large-scale $B$ mode observations of the Simons
Observatory might be used to constrain the tensor-to-scalar ratio, following on
from the forecast study in SO19. We showed that the nominal SO design may result
in a biased estimation of $r$ due to foreground contamination, which can either
be mitigated by marginalizing over a foreground residual after cleaning the maps,
or by restricting the sky area to discard the most contaminated region. This
masking removes or reduces this bias to one standard deviation or less.

In a further alternative approach, we defined a scheme to allow for spatial
variation in the parameterization of the fitted foreground model. We found that
using {\tt Healpix} pixels was unsatisfactory, as the coarse-grained pixels that
overlapped the observed regions were not fitted reliably, and could lead to large
residuals. We also defined a new spatial parameterization that split the sky into
an arbitrary number of regions following the true morphology of the spectral
behavior of the dominant foreground contaminant. With this approach, with perfect
knowledge of the spectral index variation, we found that fitting 4-6 independent
regions was sufficient to remove the dominant $r$-bias. This points to a
direction for further exploration in choosing the regions from the real data. We
found that the cross-correlation of the cleaned maps with the tracers of
synchrotron and dust provide a useful way to check for residual foregrounds that
bias the estimate for $r$. In practice, a comprehensive analysis of the real data
from SO will likely implement a set of these bias-mitigation approaches, coupled
with alternative non-parametric methods for foreground cleaning, internal null
tests to check for consistency, and tests for residual non-Gaussianity of the
maps.

\begin{acknowledgements}
	We acknowledge the use of {\tt Healpix}. BT acknowledges the support of an
	STFC studentship. DA acknowledges support from STFC through an Ernest
	Rutherford Fellowship, grant reference ST/P004474/1. MA acknowledges the
	support from the Beecroft Trust. JE acknowledges support of
	the French National Research Agency (Agence National de Recherche) grant, ANR
	B$\times$B \footnote{\url{www.bxb.space}}. This is not an
	official Simons Observatory Collaboration paper.
\end{acknowledgements}

\bibstyle{apsrev4-1}
\bibliography{library}

\begin{thebibliography}{41}%
\makeatletter
\providecommand \@ifxundefined [1]{%
 \@ifx{#1\undefined}
}%
\providecommand \@ifnum [1]{%
 \ifnum #1\expandafter \@firstoftwo
 \else \expandafter \@secondoftwo
 \fi
}%
\providecommand \@ifx [1]{%
 \ifx #1\expandafter \@firstoftwo
 \else \expandafter \@secondoftwo
 \fi
}%
\providecommand \natexlab [1]{#1}%
\providecommand \enquote  [1]{``#1''}%
\providecommand \bibnamefont  [1]{#1}%
\providecommand \bibfnamefont [1]{#1}%
\providecommand \citenamefont [1]{#1}%
\providecommand \href@noop [0]{\@secondoftwo}%
\providecommand \href [0]{\begingroup \@sanitize@url \@href}%
\providecommand \@href[1]{\@@startlink{#1}\@@href}%
\providecommand \@@href[1]{\endgroup#1\@@endlink}%
\providecommand \@sanitize@url [0]{\catcode `\\12\catcode `\$12\catcode
  `\&12\catcode `\#12\catcode `\^12\catcode `\_12\catcode `\%12\relax}%
\providecommand \@@startlink[1]{}%
\providecommand \@@endlink[0]{}%
\providecommand \url  [0]{\begingroup\@sanitize@url \@url }%
\providecommand \@url [1]{\endgroup\@href {#1}{\urlprefix }}%
\providecommand \urlprefix  [0]{URL }%
\providecommand \Eprint [0]{\href }%
\providecommand \doibase [0]{http://dx.doi.org/}%
\providecommand \selectlanguage [0]{\@gobble}%
\providecommand \bibinfo  [0]{\@secondoftwo}%
\providecommand \bibfield  [0]{\@secondoftwo}%
\providecommand \translation [1]{[#1]}%
\providecommand \BibitemOpen [0]{}%
\providecommand \bibitemStop [0]{}%
\providecommand \bibitemNoStop [0]{.\EOS\space}%
\providecommand \EOS [0]{\spacefactor3000\relax}%
\providecommand \BibitemShut  [1]{\csname bibitem#1\endcsname}%
\let\auto@bib@innerbib\@empty
\bibitem [{\citenamefont {{The Simons Observatory
  Collaboration}}(2019)}]{so_forecasts:2018}%
  \BibitemOpen
  \bibfield  {author} {\bibinfo {author} {\bibnamefont {{The Simons Observatory
  Collaboration}}},\ }\href {\doibase 10.1088/1475-7516/2019/02/056} {\bibfield
   {journal} {\bibinfo  {journal} {Journal of Cosmology and Astro-Particle
  Physics}\ }\textbf {\bibinfo {volume} {2019}},\ \bibinfo {eid} {056}
  (\bibinfo {year} {2019})},\ \Eprint {http://arxiv.org/abs/1808.07445}
  {arXiv:1808.07445 [astro-ph.CO]} \BibitemShut {NoStop}%
\bibitem [{\citenamefont {{Hui}}\ \emph {et~al.}(2018)\citenamefont {{Hui}},
  \citenamefont {{Ade}}, \citenamefont {{Ahmed}}, \citenamefont {{Aikin}},
  \citenamefont {{Alexander}}, \citenamefont {{Barkats}}, \citenamefont
  {{Benton}}, \citenamefont {{Bischoff}}, \citenamefont {{Bock}}, \citenamefont
  {{Bowens-Rubin}}, \citenamefont {{Brevik}}, \citenamefont {{Buder}},
  \citenamefont {{Bullock}}, \citenamefont {{Buza}}, \citenamefont {{Connors}},
  \citenamefont {{Cornelison}}, \citenamefont {{Crill}}, \citenamefont
  {{Crumrine}}, \citenamefont {{Dierickx}}, \citenamefont {{Duband}},
  \citenamefont {{Dvorkin}}, \citenamefont {{Filippini}}, \citenamefont
  {{Fliescher}}, \citenamefont {{Grayson}}, \citenamefont {{Hall}},
  \citenamefont {{Halpern}}, \citenamefont {{Harrison}}, \citenamefont
  {{Hildebrand t}}, \citenamefont {{Hilton}}, \citenamefont {{Irwin}},
  \citenamefont {{Kang}}, \citenamefont {{Karkare}}, \citenamefont {{Karpel}},
  \citenamefont {{Kaufman}}, \citenamefont {{Keating}}, \citenamefont
  {{Kefeli}}, \citenamefont {{Kernasovskiy}}, \citenamefont {{Kovac}},
  \citenamefont {{Kuo}}, \citenamefont {{Lau}}, \citenamefont {{Larsen}},
  \citenamefont {{Leitch}}, \citenamefont {{Lueker}}, \citenamefont
  {{Megerian}}, \citenamefont {{Moncelsi}}, \citenamefont {{Namikawa}},
  \citenamefont {{Netterfield}}, \citenamefont {{Nguyen}}, \citenamefont
  {{O'Brient}}, \citenamefont {{Ogburn}}, \citenamefont {{Palladino}},
  \citenamefont {{Pryke}}, \citenamefont {{Racine}}, \citenamefont {{Richter}},
  \citenamefont {{Schwarz}}, \citenamefont {{Schillaci}}, \citenamefont
  {{Sheehy}}, \citenamefont {{Soliman}}, \citenamefont {{St. Germaine}},
  \citenamefont {{Staniszewski}}, \citenamefont {{Steinbach}}, \citenamefont
  {{Sudiwala}}, \citenamefont {{Teply}}, \citenamefont {{Thompson}},
  \citenamefont {{Tolan}}, \citenamefont {{Tucker}}, \citenamefont {{Turner}},
  \citenamefont {{Umilt{\`a}}}, \citenamefont {{Vieregg}}, \citenamefont {{Wand
  ui}}, \citenamefont {{Weber}}, \citenamefont {{Wiebe}}, \citenamefont
  {{Willmert}}, \citenamefont {{Wong}}, \citenamefont {{Wu}}, \citenamefont
  {{Yang}}, \citenamefont {{Yoon}},\ and\ \citenamefont
  {{Zhang}}}]{hui/etal:2018}%
  \BibitemOpen
  \bibfield  {author} {\bibinfo {author} {\bibfnamefont {H.}~\bibnamefont
  {{Hui}}}, \bibinfo {author} {\bibfnamefont {P.~A.~R.}\ \bibnamefont {{Ade}}},
  \bibinfo {author} {\bibfnamefont {Z.}~\bibnamefont {{Ahmed}}}, \bibinfo
  {author} {\bibfnamefont {R.~W.}\ \bibnamefont {{Aikin}}}, \bibinfo {author}
  {\bibfnamefont {K.~D.}\ \bibnamefont {{Alexander}}}, \bibinfo {author}
  {\bibfnamefont {D.}~\bibnamefont {{Barkats}}}, \bibinfo {author}
  {\bibfnamefont {S.~J.}\ \bibnamefont {{Benton}}}, \bibinfo {author}
  {\bibfnamefont {C.~A.}\ \bibnamefont {{Bischoff}}}, \bibinfo {author}
  {\bibfnamefont {J.~J.}\ \bibnamefont {{Bock}}}, \bibinfo {author}
  {\bibfnamefont {R.}~\bibnamefont {{Bowens-Rubin}}}, \bibinfo {author}
  {\bibfnamefont {J.~A.}\ \bibnamefont {{Brevik}}}, \bibinfo {author}
  {\bibfnamefont {I.}~\bibnamefont {{Buder}}}, \bibinfo {author} {\bibfnamefont
  {E.}~\bibnamefont {{Bullock}}}, \bibinfo {author} {\bibfnamefont
  {V.}~\bibnamefont {{Buza}}}, \bibinfo {author} {\bibfnamefont
  {J.}~\bibnamefont {{Connors}}}, \bibinfo {author} {\bibfnamefont
  {J.}~\bibnamefont {{Cornelison}}}, \bibinfo {author} {\bibfnamefont {B.~P.}\
  \bibnamefont {{Crill}}}, \bibinfo {author} {\bibfnamefont {M.}~\bibnamefont
  {{Crumrine}}}, \bibinfo {author} {\bibfnamefont {M.}~\bibnamefont
  {{Dierickx}}}, \bibinfo {author} {\bibfnamefont {L.}~\bibnamefont
  {{Duband}}}, \bibinfo {author} {\bibfnamefont {C.}~\bibnamefont {{Dvorkin}}},
  \bibinfo {author} {\bibfnamefont {J.~P.}\ \bibnamefont {{Filippini}}},
  \bibinfo {author} {\bibfnamefont {S.}~\bibnamefont {{Fliescher}}}, \bibinfo
  {author} {\bibfnamefont {J.}~\bibnamefont {{Grayson}}}, \bibinfo {author}
  {\bibfnamefont {G.}~\bibnamefont {{Hall}}}, \bibinfo {author} {\bibfnamefont
  {M.}~\bibnamefont {{Halpern}}}, \bibinfo {author} {\bibfnamefont
  {S.}~\bibnamefont {{Harrison}}}, \bibinfo {author} {\bibfnamefont {S.~R.}\
  \bibnamefont {{Hildebrand t}}}, \bibinfo {author} {\bibfnamefont {G.~C.}\
  \bibnamefont {{Hilton}}}, \bibinfo {author} {\bibfnamefont {K.~D.}\
  \bibnamefont {{Irwin}}}, \bibinfo {author} {\bibfnamefont {J.}~\bibnamefont
  {{Kang}}}, \bibinfo {author} {\bibfnamefont {K.~S.}\ \bibnamefont
  {{Karkare}}}, \bibinfo {author} {\bibfnamefont {E.}~\bibnamefont {{Karpel}}},
  \bibinfo {author} {\bibfnamefont {J.~P.}\ \bibnamefont {{Kaufman}}}, \bibinfo
  {author} {\bibfnamefont {B.~G.}\ \bibnamefont {{Keating}}}, \bibinfo {author}
  {\bibfnamefont {S.}~\bibnamefont {{Kefeli}}}, \bibinfo {author}
  {\bibfnamefont {S.~A.}\ \bibnamefont {{Kernasovskiy}}}, \bibinfo {author}
  {\bibfnamefont {J.~M.}\ \bibnamefont {{Kovac}}}, \bibinfo {author}
  {\bibfnamefont {C.~L.}\ \bibnamefont {{Kuo}}}, \bibinfo {author}
  {\bibfnamefont {K.}~\bibnamefont {{Lau}}}, \bibinfo {author} {\bibfnamefont
  {N.~A.}\ \bibnamefont {{Larsen}}}, \bibinfo {author} {\bibfnamefont {E.~M.}\
  \bibnamefont {{Leitch}}}, \bibinfo {author} {\bibfnamefont {M.}~\bibnamefont
  {{Lueker}}}, \bibinfo {author} {\bibfnamefont {K.~G.}\ \bibnamefont
  {{Megerian}}}, \bibinfo {author} {\bibfnamefont {L.}~\bibnamefont
  {{Moncelsi}}}, \bibinfo {author} {\bibfnamefont {T.}~\bibnamefont
  {{Namikawa}}}, \bibinfo {author} {\bibfnamefont {C.~B.}\ \bibnamefont
  {{Netterfield}}}, \bibinfo {author} {\bibfnamefont {H.~T.}\ \bibnamefont
  {{Nguyen}}}, \bibinfo {author} {\bibfnamefont {R.}~\bibnamefont
  {{O'Brient}}}, \bibinfo {author} {\bibfnamefont {R.~W.}\ \bibnamefont
  {{Ogburn}}}, \bibinfo {author} {\bibfnamefont {S.}~\bibnamefont
  {{Palladino}}}, \bibinfo {author} {\bibfnamefont {C.}~\bibnamefont
  {{Pryke}}}, \bibinfo {author} {\bibfnamefont {B.}~\bibnamefont {{Racine}}},
  \bibinfo {author} {\bibfnamefont {S.}~\bibnamefont {{Richter}}}, \bibinfo
  {author} {\bibfnamefont {R.}~\bibnamefont {{Schwarz}}}, \bibinfo {author}
  {\bibfnamefont {A.}~\bibnamefont {{Schillaci}}}, \bibinfo {author}
  {\bibfnamefont {C.~D.}\ \bibnamefont {{Sheehy}}}, \bibinfo {author}
  {\bibfnamefont {A.}~\bibnamefont {{Soliman}}}, \bibinfo {author}
  {\bibfnamefont {T.}~\bibnamefont {{St. Germaine}}}, \bibinfo {author}
  {\bibfnamefont {Z.~K.}\ \bibnamefont {{Staniszewski}}}, \bibinfo {author}
  {\bibfnamefont {B.}~\bibnamefont {{Steinbach}}}, \bibinfo {author}
  {\bibfnamefont {R.~V.}\ \bibnamefont {{Sudiwala}}}, \bibinfo {author}
  {\bibfnamefont {G.~P.}\ \bibnamefont {{Teply}}}, \bibinfo {author}
  {\bibfnamefont {K.~L.}\ \bibnamefont {{Thompson}}}, \bibinfo {author}
  {\bibfnamefont {J.~E.}\ \bibnamefont {{Tolan}}}, \bibinfo {author}
  {\bibfnamefont {C.}~\bibnamefont {{Tucker}}}, \bibinfo {author}
  {\bibfnamefont {A.~D.}\ \bibnamefont {{Turner}}}, \bibinfo {author}
  {\bibfnamefont {C.}~\bibnamefont {{Umilt{\`a}}}}, \bibinfo {author}
  {\bibfnamefont {A.~G.}\ \bibnamefont {{Vieregg}}}, \bibinfo {author}
  {\bibfnamefont {A.}~\bibnamefont {{Wand ui}}}, \bibinfo {author}
  {\bibfnamefont {A.~C.}\ \bibnamefont {{Weber}}}, \bibinfo {author}
  {\bibfnamefont {D.~V.}\ \bibnamefont {{Wiebe}}}, \bibinfo {author}
  {\bibfnamefont {J.}~\bibnamefont {{Willmert}}}, \bibinfo {author}
  {\bibfnamefont {C.~L.}\ \bibnamefont {{Wong}}}, \bibinfo {author}
  {\bibfnamefont {W.~L.~K.}\ \bibnamefont {{Wu}}}, \bibinfo {author}
  {\bibfnamefont {E.}~\bibnamefont {{Yang}}}, \bibinfo {author} {\bibfnamefont
  {K.~W.}\ \bibnamefont {{Yoon}}}, \ and\ \bibinfo {author} {\bibfnamefont
  {C.}~\bibnamefont {{Zhang}}},\ }in\ \href {\doibase 10.1117/12.2311725}
  {\emph {\bibinfo {booktitle} {Millimeter, Submillimeter, and Far-Infrared
  Detectors and Instrumentation for Astronomy IX}}},\ \bibinfo {series}
  {Society of Photo-Optical Instrumentation Engineers (SPIE) Conference
  Series}, Vol.\ \bibinfo {volume} {10708}\ (\bibinfo {year} {2018})\ p.\
  \bibinfo {pages} {1070807},\ \Eprint {http://arxiv.org/abs/1808.00568}
  {arXiv:1808.00568 [astro-ph.IM]} \BibitemShut {NoStop}%
\bibitem [{\citenamefont {{Planck Collaboration}}(2018)}]{planck_cosmo:2018}%
  \BibitemOpen
  \bibfield  {author} {\bibinfo {author} {\bibnamefont {{Planck
  Collaboration}}},\ }\href@noop {} {\bibfield  {journal} {\bibinfo  {journal}
  {arXiv e-prints}\ ,\ \bibinfo {eid} {arXiv:1807.06209}} (\bibinfo {year}
  {2018})},\ \Eprint {http://arxiv.org/abs/1807.06209} {arXiv:1807.06209
  [astro-ph.CO]} \BibitemShut {NoStop}%
\bibitem [{\citenamefont {Kamionkowski}\ \emph {et~al.}(1997)\citenamefont
  {Kamionkowski}, \citenamefont {Kosowsky},\ and\ \citenamefont
  {Stebbins}}]{kamkionkowski/etal:1997}%
  \BibitemOpen
  \bibfield  {author} {\bibinfo {author} {\bibfnamefont {M.}~\bibnamefont
  {Kamionkowski}}, \bibinfo {author} {\bibfnamefont {A.}~\bibnamefont
  {Kosowsky}}, \ and\ \bibinfo {author} {\bibfnamefont {A.}~\bibnamefont
  {Stebbins}},\ }\href {\doibase 10.1103/PhysRevD.55.7368} {\bibfield
  {journal} {\bibinfo  {journal} {Phys. Rev. D}\ }\textbf {\bibinfo {volume}
  {55}},\ \bibinfo {pages} {7368} (\bibinfo {year} {1997})}\BibitemShut
  {NoStop}%
\bibitem [{\citenamefont {{Seljak}}\ and\ \citenamefont
  {{Zaldarriaga}}(1996)}]{seljak/zaldarriaga:1996}%
  \BibitemOpen
  \bibfield  {author} {\bibinfo {author} {\bibfnamefont {U.}~\bibnamefont
  {{Seljak}}}\ and\ \bibinfo {author} {\bibfnamefont {M.}~\bibnamefont
  {{Zaldarriaga}}},\ }\href {\doibase 10.1086/177793} {\bibfield  {journal}
  {\bibinfo  {journal} {\apj}\ }\textbf {\bibinfo {volume} {469}},\ \bibinfo
  {pages} {437} (\bibinfo {year} {1996})},\ \Eprint
  {http://arxiv.org/abs/astro-ph/9603033} {arXiv:astro-ph/9603033 [astro-ph]}
  \BibitemShut {NoStop}%
\bibitem [{\citenamefont {{Ijjas}}\ and\ \citenamefont
  {{Steinhardt}}(2018)}]{ijjas/steinhardt:2018}%
  \BibitemOpen
  \bibfield  {author} {\bibinfo {author} {\bibfnamefont {A.}~\bibnamefont
  {{Ijjas}}}\ and\ \bibinfo {author} {\bibfnamefont {P.~J.}\ \bibnamefont
  {{Steinhardt}}},\ }\href {\doibase 10.1088/1361-6382/aac482} {\bibfield
  {journal} {\bibinfo  {journal} {Classical and Quantum Gravity}\ }\textbf
  {\bibinfo {volume} {35}},\ \bibinfo {eid} {135004} (\bibinfo {year}
  {2018})},\ \Eprint {http://arxiv.org/abs/1803.01961} {arXiv:1803.01961}
  \BibitemShut {NoStop}%
\bibitem [{\citenamefont {{BICEP2 Collaboration}}\ and\ \citenamefont {{Keck
  Array Collaboration}}(2018)}]{bicep2/keck:2018}%
  \BibitemOpen
  \bibfield  {author} {\bibinfo {author} {\bibnamefont {{BICEP2
  Collaboration}}}\ and\ \bibinfo {author} {\bibnamefont {{Keck Array
  Collaboration}}},\ }\href {\doibase 10.1103/PhysRevLett.121.221301}
  {\bibfield  {journal} {\bibinfo  {journal} {Physical Review Letters}\
  }\textbf {\bibinfo {volume} {121}},\ \bibinfo {eid} {221301} (\bibinfo {year}
  {2018})},\ \Eprint {http://arxiv.org/abs/1810.05216} {arXiv:1810.05216}
  \BibitemShut {NoStop}%
\bibitem [{\citenamefont {{Thornton}}\ \emph {et~al.}(2016)\citenamefont
  {{Thornton}}, \citenamefont {{Ade}}, \citenamefont {{Aiola}}, \citenamefont
  {{Angil{\`e}}}, \citenamefont {{Amiri}}, \citenamefont {{Beall}},
  \citenamefont {{Becker}}, \citenamefont {{Cho}}, \citenamefont {{Choi}},
  \citenamefont {{Corlies}}, \citenamefont {{Coughlin}}, \citenamefont
  {{Datta}}, \citenamefont {{Devlin}}, \citenamefont {{Dicker}}, \citenamefont
  {{D{\"u}nner}}, \citenamefont {{Fowler}}, \citenamefont {{Fox}},
  \citenamefont {{Gallardo}}, \citenamefont {{Gao}}, \citenamefont {{Grace}},
  \citenamefont {{Halpern}}, \citenamefont {{Hasselfield}}, \citenamefont
  {{Henderson}}, \citenamefont {{Hilton}}, \citenamefont {{Hincks}},
  \citenamefont {{Ho}}, \citenamefont {{Hubmayr}}, \citenamefont {{Irwin}},
  \citenamefont {{Klein}}, \citenamefont {{Koopman}}, \citenamefont {{Li}},
  \citenamefont {{Louis}}, \citenamefont {{Lungu}}, \citenamefont {{Maurin}},
  \citenamefont {{McMahon}}, \citenamefont {{Munson}}, \citenamefont {{Naess}},
  \citenamefont {{Nati}}, \citenamefont {{Newburgh}}, \citenamefont
  {{Nibarger}}, \citenamefont {{Niemack}}, \citenamefont {{Niraula}},
  \citenamefont {{Nolta}}, \citenamefont {{Page}}, \citenamefont {{Pappas}},
  \citenamefont {{Schillaci}}, \citenamefont {{Schmitt}}, \citenamefont
  {{Sehgal}}, \citenamefont {{Sievers}}, \citenamefont {{Simon}}, \citenamefont
  {{Staggs}}, \citenamefont {{Tucker}}, \citenamefont {{Uehara}}, \citenamefont
  {{van Lanen}}, \citenamefont {{Ward}},\ and\ \citenamefont
  {{Wollack}}}]{actpol_instrument:2016}%
  \BibitemOpen
  \bibfield  {author} {\bibinfo {author} {\bibfnamefont {R.~J.}\ \bibnamefont
  {{Thornton}}}, \bibinfo {author} {\bibfnamefont {P.~A.~R.}\ \bibnamefont
  {{Ade}}}, \bibinfo {author} {\bibfnamefont {S.}~\bibnamefont {{Aiola}}},
  \bibinfo {author} {\bibfnamefont {F.~E.}\ \bibnamefont {{Angil{\`e}}}},
  \bibinfo {author} {\bibfnamefont {M.}~\bibnamefont {{Amiri}}}, \bibinfo
  {author} {\bibfnamefont {J.~A.}\ \bibnamefont {{Beall}}}, \bibinfo {author}
  {\bibfnamefont {D.~T.}\ \bibnamefont {{Becker}}}, \bibinfo {author}
  {\bibfnamefont {H.-M.}\ \bibnamefont {{Cho}}}, \bibinfo {author}
  {\bibfnamefont {S.~K.}\ \bibnamefont {{Choi}}}, \bibinfo {author}
  {\bibfnamefont {P.}~\bibnamefont {{Corlies}}}, \bibinfo {author}
  {\bibfnamefont {K.~P.}\ \bibnamefont {{Coughlin}}}, \bibinfo {author}
  {\bibfnamefont {R.}~\bibnamefont {{Datta}}}, \bibinfo {author} {\bibfnamefont
  {M.~J.}\ \bibnamefont {{Devlin}}}, \bibinfo {author} {\bibfnamefont {S.~R.}\
  \bibnamefont {{Dicker}}}, \bibinfo {author} {\bibfnamefont {R.}~\bibnamefont
  {{D{\"u}nner}}}, \bibinfo {author} {\bibfnamefont {J.~W.}\ \bibnamefont
  {{Fowler}}}, \bibinfo {author} {\bibfnamefont {A.~E.}\ \bibnamefont {{Fox}}},
  \bibinfo {author} {\bibfnamefont {P.~A.}\ \bibnamefont {{Gallardo}}},
  \bibinfo {author} {\bibfnamefont {J.}~\bibnamefont {{Gao}}}, \bibinfo
  {author} {\bibfnamefont {E.}~\bibnamefont {{Grace}}}, \bibinfo {author}
  {\bibfnamefont {M.}~\bibnamefont {{Halpern}}}, \bibinfo {author}
  {\bibfnamefont {M.}~\bibnamefont {{Hasselfield}}}, \bibinfo {author}
  {\bibfnamefont {S.~W.}\ \bibnamefont {{Henderson}}}, \bibinfo {author}
  {\bibfnamefont {G.~C.}\ \bibnamefont {{Hilton}}}, \bibinfo {author}
  {\bibfnamefont {A.~D.}\ \bibnamefont {{Hincks}}}, \bibinfo {author}
  {\bibfnamefont {S.~P.}\ \bibnamefont {{Ho}}}, \bibinfo {author}
  {\bibfnamefont {J.}~\bibnamefont {{Hubmayr}}}, \bibinfo {author}
  {\bibfnamefont {K.~D.}\ \bibnamefont {{Irwin}}}, \bibinfo {author}
  {\bibfnamefont {J.}~\bibnamefont {{Klein}}}, \bibinfo {author} {\bibfnamefont
  {B.}~\bibnamefont {{Koopman}}}, \bibinfo {author} {\bibfnamefont
  {D.}~\bibnamefont {{Li}}}, \bibinfo {author} {\bibfnamefont {T.}~\bibnamefont
  {{Louis}}}, \bibinfo {author} {\bibfnamefont {M.}~\bibnamefont {{Lungu}}},
  \bibinfo {author} {\bibfnamefont {L.}~\bibnamefont {{Maurin}}}, \bibinfo
  {author} {\bibfnamefont {J.}~\bibnamefont {{McMahon}}}, \bibinfo {author}
  {\bibfnamefont {C.~D.}\ \bibnamefont {{Munson}}}, \bibinfo {author}
  {\bibfnamefont {S.}~\bibnamefont {{Naess}}}, \bibinfo {author} {\bibfnamefont
  {F.}~\bibnamefont {{Nati}}}, \bibinfo {author} {\bibfnamefont
  {L.}~\bibnamefont {{Newburgh}}}, \bibinfo {author} {\bibfnamefont
  {J.}~\bibnamefont {{Nibarger}}}, \bibinfo {author} {\bibfnamefont {M.~D.}\
  \bibnamefont {{Niemack}}}, \bibinfo {author} {\bibfnamefont {P.}~\bibnamefont
  {{Niraula}}}, \bibinfo {author} {\bibfnamefont {M.~R.}\ \bibnamefont
  {{Nolta}}}, \bibinfo {author} {\bibfnamefont {L.~A.}\ \bibnamefont {{Page}}},
  \bibinfo {author} {\bibfnamefont {C.~G.}\ \bibnamefont {{Pappas}}}, \bibinfo
  {author} {\bibfnamefont {A.}~\bibnamefont {{Schillaci}}}, \bibinfo {author}
  {\bibfnamefont {B.~L.}\ \bibnamefont {{Schmitt}}}, \bibinfo {author}
  {\bibfnamefont {N.}~\bibnamefont {{Sehgal}}}, \bibinfo {author}
  {\bibfnamefont {J.~L.}\ \bibnamefont {{Sievers}}}, \bibinfo {author}
  {\bibfnamefont {S.~M.}\ \bibnamefont {{Simon}}}, \bibinfo {author}
  {\bibfnamefont {S.~T.}\ \bibnamefont {{Staggs}}}, \bibinfo {author}
  {\bibfnamefont {C.}~\bibnamefont {{Tucker}}}, \bibinfo {author}
  {\bibfnamefont {M.}~\bibnamefont {{Uehara}}}, \bibinfo {author}
  {\bibfnamefont {J.}~\bibnamefont {{van Lanen}}}, \bibinfo {author}
  {\bibfnamefont {J.~T.}\ \bibnamefont {{Ward}}}, \ and\ \bibinfo {author}
  {\bibfnamefont {E.~J.}\ \bibnamefont {{Wollack}}},\ }\href {\doibase
  10.3847/1538-4365/227/2/21} {\bibfield  {journal} {\bibinfo  {journal}
  {\apjs}\ }\textbf {\bibinfo {volume} {227}},\ \bibinfo {eid} {21} (\bibinfo
  {year} {2016})},\ \Eprint {http://arxiv.org/abs/1605.06569} {arXiv:1605.06569
  [astro-ph.IM]} \BibitemShut {NoStop}%
\bibitem [{\citenamefont {{Benson}}\ \emph {et~al.}(2014)\citenamefont
  {{Benson}}, \citenamefont {{Ade}}, \citenamefont {{Ahmed}}, \citenamefont
  {{Allen}}, \citenamefont {{Arnold}}, \citenamefont {{Austermann}},
  \citenamefont {{Bender}}, \citenamefont {{Bleem}}, \citenamefont
  {{Carlstrom}}, \citenamefont {{Chang}}, \citenamefont {{Cho}}, \citenamefont
  {{Cliche}}, \citenamefont {{Crawford}}, \citenamefont {{Cukierman}},
  \citenamefont {{de Haan}}, \citenamefont {{Dobbs}}, \citenamefont
  {{Dutcher}}, \citenamefont {{Everett}}, \citenamefont {{Gilbert}},
  \citenamefont {{Halverson}}, \citenamefont {{Hanson}}, \citenamefont
  {{Harrington}}, \citenamefont {{Hattori}}, \citenamefont {{Henning}},
  \citenamefont {{Hilton}}, \citenamefont {{Holder}}, \citenamefont
  {{Holzapfel}}, \citenamefont {{Irwin}}, \citenamefont {{Keisler}},
  \citenamefont {{Knox}}, \citenamefont {{Kubik}}, \citenamefont {{Kuo}},
  \citenamefont {{Lee}}, \citenamefont {{Leitch}}, \citenamefont {{Li}},
  \citenamefont {{McDonald}}, \citenamefont {{Meyer}}, \citenamefont
  {{Montgomery}}, \citenamefont {{Myers}}, \citenamefont {{Natoli}},
  \citenamefont {{Nguyen}}, \citenamefont {{Novosad}}, \citenamefont {{Padin}},
  \citenamefont {{Pan}}, \citenamefont {{Pearson}}, \citenamefont
  {{Reichardt}}, \citenamefont {{Ruhl}}, \citenamefont {{Saliwanchik}},
  \citenamefont {{Simard}}, \citenamefont {{Smecher}}, \citenamefont {{Sayre}},
  \citenamefont {{Shirokoff}}, \citenamefont {{Stark}}, \citenamefont
  {{Story}}, \citenamefont {{Suzuki}}, \citenamefont {{Thompson}},
  \citenamefont {{Tucker}}, \citenamefont {{Vanderlinde}}, \citenamefont
  {{Vieira}}, \citenamefont {{Vikhlinin}}, \citenamefont {{Wang}},
  \citenamefont {{Yefremenko}},\ and\ \citenamefont {{Yoon}}}]{sptpol3g}%
  \BibitemOpen
  \bibfield  {author} {\bibinfo {author} {\bibfnamefont {B.~A.}\ \bibnamefont
  {{Benson}}}, \bibinfo {author} {\bibfnamefont {P.~A.~R.}\ \bibnamefont
  {{Ade}}}, \bibinfo {author} {\bibfnamefont {Z.}~\bibnamefont {{Ahmed}}},
  \bibinfo {author} {\bibfnamefont {S.~W.}\ \bibnamefont {{Allen}}}, \bibinfo
  {author} {\bibfnamefont {K.}~\bibnamefont {{Arnold}}}, \bibinfo {author}
  {\bibfnamefont {J.~E.}\ \bibnamefont {{Austermann}}}, \bibinfo {author}
  {\bibfnamefont {A.~N.}\ \bibnamefont {{Bender}}}, \bibinfo {author}
  {\bibfnamefont {L.~E.}\ \bibnamefont {{Bleem}}}, \bibinfo {author}
  {\bibfnamefont {J.~E.}\ \bibnamefont {{Carlstrom}}}, \bibinfo {author}
  {\bibfnamefont {C.~L.}\ \bibnamefont {{Chang}}}, \bibinfo {author}
  {\bibfnamefont {H.~M.}\ \bibnamefont {{Cho}}}, \bibinfo {author}
  {\bibfnamefont {J.~F.}\ \bibnamefont {{Cliche}}}, \bibinfo {author}
  {\bibfnamefont {T.~M.}\ \bibnamefont {{Crawford}}}, \bibinfo {author}
  {\bibfnamefont {A.}~\bibnamefont {{Cukierman}}}, \bibinfo {author}
  {\bibfnamefont {T.}~\bibnamefont {{de Haan}}}, \bibinfo {author}
  {\bibfnamefont {M.~A.}\ \bibnamefont {{Dobbs}}}, \bibinfo {author}
  {\bibfnamefont {D.}~\bibnamefont {{Dutcher}}}, \bibinfo {author}
  {\bibfnamefont {W.}~\bibnamefont {{Everett}}}, \bibinfo {author}
  {\bibfnamefont {A.}~\bibnamefont {{Gilbert}}}, \bibinfo {author}
  {\bibfnamefont {N.~W.}\ \bibnamefont {{Halverson}}}, \bibinfo {author}
  {\bibfnamefont {D.}~\bibnamefont {{Hanson}}}, \bibinfo {author}
  {\bibfnamefont {N.~L.}\ \bibnamefont {{Harrington}}}, \bibinfo {author}
  {\bibfnamefont {K.}~\bibnamefont {{Hattori}}}, \bibinfo {author}
  {\bibfnamefont {J.~W.}\ \bibnamefont {{Henning}}}, \bibinfo {author}
  {\bibfnamefont {G.~C.}\ \bibnamefont {{Hilton}}}, \bibinfo {author}
  {\bibfnamefont {G.~P.}\ \bibnamefont {{Holder}}}, \bibinfo {author}
  {\bibfnamefont {W.~L.}\ \bibnamefont {{Holzapfel}}}, \bibinfo {author}
  {\bibfnamefont {K.~D.}\ \bibnamefont {{Irwin}}}, \bibinfo {author}
  {\bibfnamefont {R.}~\bibnamefont {{Keisler}}}, \bibinfo {author}
  {\bibfnamefont {L.}~\bibnamefont {{Knox}}}, \bibinfo {author} {\bibfnamefont
  {D.}~\bibnamefont {{Kubik}}}, \bibinfo {author} {\bibfnamefont {C.~L.}\
  \bibnamefont {{Kuo}}}, \bibinfo {author} {\bibfnamefont {A.~T.}\ \bibnamefont
  {{Lee}}}, \bibinfo {author} {\bibfnamefont {E.~M.}\ \bibnamefont {{Leitch}}},
  \bibinfo {author} {\bibfnamefont {D.}~\bibnamefont {{Li}}}, \bibinfo {author}
  {\bibfnamefont {M.}~\bibnamefont {{McDonald}}}, \bibinfo {author}
  {\bibfnamefont {S.~S.}\ \bibnamefont {{Meyer}}}, \bibinfo {author}
  {\bibfnamefont {J.}~\bibnamefont {{Montgomery}}}, \bibinfo {author}
  {\bibfnamefont {M.}~\bibnamefont {{Myers}}}, \bibinfo {author} {\bibfnamefont
  {T.}~\bibnamefont {{Natoli}}}, \bibinfo {author} {\bibfnamefont
  {H.}~\bibnamefont {{Nguyen}}}, \bibinfo {author} {\bibfnamefont
  {V.}~\bibnamefont {{Novosad}}}, \bibinfo {author} {\bibfnamefont
  {S.}~\bibnamefont {{Padin}}}, \bibinfo {author} {\bibfnamefont
  {Z.}~\bibnamefont {{Pan}}}, \bibinfo {author} {\bibfnamefont
  {J.}~\bibnamefont {{Pearson}}}, \bibinfo {author} {\bibfnamefont
  {C.}~\bibnamefont {{Reichardt}}}, \bibinfo {author} {\bibfnamefont {J.~E.}\
  \bibnamefont {{Ruhl}}}, \bibinfo {author} {\bibfnamefont {B.~R.}\
  \bibnamefont {{Saliwanchik}}}, \bibinfo {author} {\bibfnamefont
  {G.}~\bibnamefont {{Simard}}}, \bibinfo {author} {\bibfnamefont
  {G.}~\bibnamefont {{Smecher}}}, \bibinfo {author} {\bibfnamefont {J.~T.}\
  \bibnamefont {{Sayre}}}, \bibinfo {author} {\bibfnamefont {E.}~\bibnamefont
  {{Shirokoff}}}, \bibinfo {author} {\bibfnamefont {A.~A.}\ \bibnamefont
  {{Stark}}}, \bibinfo {author} {\bibfnamefont {K.}~\bibnamefont {{Story}}},
  \bibinfo {author} {\bibfnamefont {A.}~\bibnamefont {{Suzuki}}}, \bibinfo
  {author} {\bibfnamefont {K.~L.}\ \bibnamefont {{Thompson}}}, \bibinfo
  {author} {\bibfnamefont {C.}~\bibnamefont {{Tucker}}}, \bibinfo {author}
  {\bibfnamefont {K.}~\bibnamefont {{Vanderlinde}}}, \bibinfo {author}
  {\bibfnamefont {J.~D.}\ \bibnamefont {{Vieira}}}, \bibinfo {author}
  {\bibfnamefont {A.}~\bibnamefont {{Vikhlinin}}}, \bibinfo {author}
  {\bibfnamefont {G.}~\bibnamefont {{Wang}}}, \bibinfo {author} {\bibfnamefont
  {V.}~\bibnamefont {{Yefremenko}}}, \ and\ \bibinfo {author} {\bibfnamefont
  {K.~W.}\ \bibnamefont {{Yoon}}},\ }in\ \href {\doibase 10.1117/12.2057305}
  {\emph {\bibinfo {booktitle} {Millimeter, Submillimeter, and Far-Infrared
  Detectors and Instrumentation for Astronomy VII}}},\ \bibinfo {series}
  {\procspie}, Vol.\ \bibinfo {volume} {9153}\ (\bibinfo {year} {2014})\ p.\
  \bibinfo {pages} {91531P},\ \Eprint {http://arxiv.org/abs/1407.2973}
  {arXiv:1407.2973 [astro-ph.IM]} \BibitemShut {NoStop}%
\bibitem [{\citenamefont {{Kang}}\ \emph {et~al.}(2018)\citenamefont {{Kang}},
  \citenamefont {{Ade}}, \citenamefont {{Ahmed}}, \citenamefont {{Aikin}},
  \citenamefont {{Alexander}}, \citenamefont {{Barkats}}, \citenamefont
  {{Benton}}, \citenamefont {{Bischoff}}, \citenamefont {{Bock}}, \citenamefont
  {{Boenish}}, \citenamefont {{Bowens-Rubin}}, \citenamefont {{Brevik}},
  \citenamefont {{Buder}}, \citenamefont {{Bullock}}, \citenamefont {{Buza}},
  \citenamefont {{Connors}}, \citenamefont {{Cornelison}}, \citenamefont
  {{Crill}}, \citenamefont {{Crumrine}}, \citenamefont {{Dierickx}},
  \citenamefont {{Duband}}, \citenamefont {{Dvorkin}}, \citenamefont
  {{Filippini}}, \citenamefont {{Fliescher}}, \citenamefont {{Grayson}},
  \citenamefont {{Hall}}, \citenamefont {{Halpern}}, \citenamefont
  {{Harrison}}, \citenamefont {{Hildebrandt}}, \citenamefont {{Hilton}},
  \citenamefont {{Hui}}, \citenamefont {{Irwin}}, \citenamefont {{Karkare}},
  \citenamefont {{Karpel}}, \citenamefont {{Kaufman}}, \citenamefont
  {{Keating}}, \citenamefont {{Kefeli}}, \citenamefont {{Kernasovskiy}},
  \citenamefont {{Kovac}}, \citenamefont {{Kuo}}, \citenamefont {{Larsen}},
  \citenamefont {{Lau}}, \citenamefont {{Leitch}}, \citenamefont {{Lueker}},
  \citenamefont {{Megerian}}, \citenamefont {{Moncelsi}}, \citenamefont
  {{Namikawa}}, \citenamefont {{Netterfield}}, \citenamefont {{Nguyen}},
  \citenamefont {{O'Brient}}, \citenamefont {{Ogburn}}, \citenamefont
  {{Palladino}}, \citenamefont {{Pryke}}, \citenamefont {{Racine}},
  \citenamefont {{Richter}}, \citenamefont {{Schillaci}}, \citenamefont
  {{Schwarz}}, \citenamefont {{Sheehy}}, \citenamefont {{Soliman}},
  \citenamefont {{St. Germaine}}, \citenamefont {{Staniszewski}}, \citenamefont
  {{Steinbach}}, \citenamefont {{Sudiwala}}, \citenamefont {{Teply}},
  \citenamefont {{Thompson}}, \citenamefont {{Tolan}}, \citenamefont
  {{Tucker}}, \citenamefont {{Turner}}, \citenamefont {{Umilt{\`a}}},
  \citenamefont {{Vieregg}}, \citenamefont {{Wandui}}, \citenamefont {{Weber}},
  \citenamefont {{Wiebe}}, \citenamefont {{Willmert}}, \citenamefont {{Wong}},
  \citenamefont {{Wu}}, \citenamefont {{Yang}}, \citenamefont {{Yoon}},\ and\
  \citenamefont {{Zhang}}}]{bicep3:2017}%
  \BibitemOpen
  \bibfield  {author} {\bibinfo {author} {\bibfnamefont {J.~H.}\ \bibnamefont
  {{Kang}}}, \bibinfo {author} {\bibfnamefont {P.~A.~R.}\ \bibnamefont
  {{Ade}}}, \bibinfo {author} {\bibfnamefont {Z.}~\bibnamefont {{Ahmed}}},
  \bibinfo {author} {\bibfnamefont {R.~W.}\ \bibnamefont {{Aikin}}}, \bibinfo
  {author} {\bibfnamefont {K.~D.}\ \bibnamefont {{Alexander}}}, \bibinfo
  {author} {\bibfnamefont {D.}~\bibnamefont {{Barkats}}}, \bibinfo {author}
  {\bibfnamefont {S.~J.}\ \bibnamefont {{Benton}}}, \bibinfo {author}
  {\bibfnamefont {C.~A.}\ \bibnamefont {{Bischoff}}}, \bibinfo {author}
  {\bibfnamefont {J.~J.}\ \bibnamefont {{Bock}}}, \bibinfo {author}
  {\bibfnamefont {H.}~\bibnamefont {{Boenish}}}, \bibinfo {author}
  {\bibfnamefont {R.}~\bibnamefont {{Bowens-Rubin}}}, \bibinfo {author}
  {\bibfnamefont {J.~A.}\ \bibnamefont {{Brevik}}}, \bibinfo {author}
  {\bibfnamefont {I.}~\bibnamefont {{Buder}}}, \bibinfo {author} {\bibfnamefont
  {E.}~\bibnamefont {{Bullock}}}, \bibinfo {author} {\bibfnamefont
  {V.}~\bibnamefont {{Buza}}}, \bibinfo {author} {\bibfnamefont
  {J.}~\bibnamefont {{Connors}}}, \bibinfo {author} {\bibfnamefont
  {J.}~\bibnamefont {{Cornelison}}}, \bibinfo {author} {\bibfnamefont {B.~P.}\
  \bibnamefont {{Crill}}}, \bibinfo {author} {\bibfnamefont {M.}~\bibnamefont
  {{Crumrine}}}, \bibinfo {author} {\bibfnamefont {M.}~\bibnamefont
  {{Dierickx}}}, \bibinfo {author} {\bibfnamefont {L.}~\bibnamefont
  {{Duband}}}, \bibinfo {author} {\bibfnamefont {C.}~\bibnamefont {{Dvorkin}}},
  \bibinfo {author} {\bibfnamefont {J.~P.}\ \bibnamefont {{Filippini}}},
  \bibinfo {author} {\bibfnamefont {S.}~\bibnamefont {{Fliescher}}}, \bibinfo
  {author} {\bibfnamefont {J.~A.}\ \bibnamefont {{Grayson}}}, \bibinfo {author}
  {\bibfnamefont {G.}~\bibnamefont {{Hall}}}, \bibinfo {author} {\bibfnamefont
  {M.}~\bibnamefont {{Halpern}}}, \bibinfo {author} {\bibfnamefont
  {S.}~\bibnamefont {{Harrison}}}, \bibinfo {author} {\bibfnamefont {S.~R.}\
  \bibnamefont {{Hildebrandt}}}, \bibinfo {author} {\bibfnamefont {G.~C.}\
  \bibnamefont {{Hilton}}}, \bibinfo {author} {\bibfnamefont {H.}~\bibnamefont
  {{Hui}}}, \bibinfo {author} {\bibfnamefont {K.~D.}\ \bibnamefont {{Irwin}}},
  \bibinfo {author} {\bibfnamefont {K.~S.}\ \bibnamefont {{Karkare}}}, \bibinfo
  {author} {\bibfnamefont {E.}~\bibnamefont {{Karpel}}}, \bibinfo {author}
  {\bibfnamefont {J.~P.}\ \bibnamefont {{Kaufman}}}, \bibinfo {author}
  {\bibfnamefont {B.~G.}\ \bibnamefont {{Keating}}}, \bibinfo {author}
  {\bibfnamefont {S.}~\bibnamefont {{Kefeli}}}, \bibinfo {author}
  {\bibfnamefont {S.~A.}\ \bibnamefont {{Kernasovskiy}}}, \bibinfo {author}
  {\bibfnamefont {J.~M.}\ \bibnamefont {{Kovac}}}, \bibinfo {author}
  {\bibfnamefont {C.~L.}\ \bibnamefont {{Kuo}}}, \bibinfo {author}
  {\bibfnamefont {N.~A.}\ \bibnamefont {{Larsen}}}, \bibinfo {author}
  {\bibfnamefont {K.}~\bibnamefont {{Lau}}}, \bibinfo {author} {\bibfnamefont
  {E.~M.}\ \bibnamefont {{Leitch}}}, \bibinfo {author} {\bibfnamefont
  {M.}~\bibnamefont {{Lueker}}}, \bibinfo {author} {\bibfnamefont {K.~G.}\
  \bibnamefont {{Megerian}}}, \bibinfo {author} {\bibfnamefont
  {L.}~\bibnamefont {{Moncelsi}}}, \bibinfo {author} {\bibfnamefont
  {T.}~\bibnamefont {{Namikawa}}}, \bibinfo {author} {\bibfnamefont
  {B.}~\bibnamefont {{Netterfield}}}, \bibinfo {author} {\bibfnamefont {H.~T.}\
  \bibnamefont {{Nguyen}}}, \bibinfo {author} {\bibfnamefont {R.}~\bibnamefont
  {{O'Brient}}}, \bibinfo {author} {\bibfnamefont {R.~W.}\ \bibnamefont
  {{Ogburn}}}, \bibinfo {author} {\bibfnamefont {S.}~\bibnamefont
  {{Palladino}}}, \bibinfo {author} {\bibfnamefont {C.}~\bibnamefont
  {{Pryke}}}, \bibinfo {author} {\bibfnamefont {B.}~\bibnamefont {{Racine}}},
  \bibinfo {author} {\bibfnamefont {S.}~\bibnamefont {{Richter}}}, \bibinfo
  {author} {\bibfnamefont {A.}~\bibnamefont {{Schillaci}}}, \bibinfo {author}
  {\bibfnamefont {R.}~\bibnamefont {{Schwarz}}}, \bibinfo {author}
  {\bibfnamefont {C.~D.}\ \bibnamefont {{Sheehy}}}, \bibinfo {author}
  {\bibfnamefont {A.}~\bibnamefont {{Soliman}}}, \bibinfo {author}
  {\bibfnamefont {T.}~\bibnamefont {{St. Germaine}}}, \bibinfo {author}
  {\bibfnamefont {Z.~K.}\ \bibnamefont {{Staniszewski}}}, \bibinfo {author}
  {\bibfnamefont {B.}~\bibnamefont {{Steinbach}}}, \bibinfo {author}
  {\bibfnamefont {R.~V.}\ \bibnamefont {{Sudiwala}}}, \bibinfo {author}
  {\bibfnamefont {G.~P.}\ \bibnamefont {{Teply}}}, \bibinfo {author}
  {\bibfnamefont {K.~L.}\ \bibnamefont {{Thompson}}}, \bibinfo {author}
  {\bibfnamefont {J.~E.}\ \bibnamefont {{Tolan}}}, \bibinfo {author}
  {\bibfnamefont {C.}~\bibnamefont {{Tucker}}}, \bibinfo {author}
  {\bibfnamefont {A.~D.}\ \bibnamefont {{Turner}}}, \bibinfo {author}
  {\bibfnamefont {C.}~\bibnamefont {{Umilt{\`a}}}}, \bibinfo {author}
  {\bibfnamefont {A.~G.}\ \bibnamefont {{Vieregg}}}, \bibinfo {author}
  {\bibfnamefont {A.}~\bibnamefont {{Wandui}}}, \bibinfo {author}
  {\bibfnamefont {A.~C.}\ \bibnamefont {{Weber}}}, \bibinfo {author}
  {\bibfnamefont {D.~V.}\ \bibnamefont {{Wiebe}}}, \bibinfo {author}
  {\bibfnamefont {J.}~\bibnamefont {{Willmert}}}, \bibinfo {author}
  {\bibfnamefont {C.~L.}\ \bibnamefont {{Wong}}}, \bibinfo {author}
  {\bibfnamefont {W.~L.~K.}\ \bibnamefont {{Wu}}}, \bibinfo {author}
  {\bibfnamefont {H.}~\bibnamefont {{Yang}}}, \bibinfo {author} {\bibfnamefont
  {W.}~\bibnamefont {{Yoon}}, \bibfnamefont {K.}}, \ and\ \bibinfo {author}
  {\bibfnamefont {C.}~\bibnamefont {{Zhang}}},\ }in\ \href {\doibase
  10.1117/12.2313854} {\emph {\bibinfo {booktitle} {Millimeter, Submillimeter,
  and Far-Infrared Detectors and Instrumentation for Astronomy IX}}},\ \bibinfo
  {series} {Society of Photo-Optical Instrumentation Engineers (SPIE)
  Conference Series}, Vol.\ \bibinfo {volume} {10708}\ (\bibinfo {year}
  {2018})\ p.\ \bibinfo {pages} {107082N},\ \Eprint
  {http://arxiv.org/abs/1808.00567} {arXiv:1808.00567 [astro-ph.IM]}
  \BibitemShut {NoStop}%
\bibitem [{\citenamefont {{Suzuki}}\ \emph {et~al.}(2016)\citenamefont
  {{Suzuki}}, \citenamefont {{Ade}}, \citenamefont {{Akiba}}, \citenamefont
  {{Aleman}}, \citenamefont {{Arnold}}, \citenamefont {{Baccigalupi}},
  \citenamefont {{Barch}}, \citenamefont {{Barron}}, \citenamefont {{Bender}},
  \citenamefont {{Boettger}}, \citenamefont {{Borrill}}, \citenamefont
  {{Chapman}}, \citenamefont {{Chinone}}, \citenamefont {{Cukierman}},
  \citenamefont {{Dobbs}}, \citenamefont {{Ducout}}, \citenamefont {{Dunner}},
  \citenamefont {{Elleflot}}, \citenamefont {{Errard}}, \citenamefont
  {{Fabbian}}, \citenamefont {{Feeney}}, \citenamefont {{Feng}}, \citenamefont
  {{Fujino}}, \citenamefont {{Fuller}}, \citenamefont {{Gilbert}},
  \citenamefont {{Goeckner-Wald}}, \citenamefont {{Groh}}, \citenamefont
  {{Haan}}, \citenamefont {{Hall}}, \citenamefont {{Halverson}}, \citenamefont
  {{Hamada}}, \citenamefont {{Hasegawa}}, \citenamefont {{Hattori}},
  \citenamefont {{Hazumi}}, \citenamefont {{Hill}}, \citenamefont
  {{Holzapfel}}, \citenamefont {{Hori}}, \citenamefont {{Howe}}, \citenamefont
  {{Inoue}}, \citenamefont {{Irie}}, \citenamefont {{Jaehnig}}, \citenamefont
  {{Jaffe}}, \citenamefont {{Jeong}}, \citenamefont {{Katayama}}, \citenamefont
  {{Kaufman}}, \citenamefont {{Kazemzadeh}}, \citenamefont {{Keating}},
  \citenamefont {{Kermish}}, \citenamefont {{Keskitalo}}, \citenamefont
  {{Kisner}}, \citenamefont {{Kusaka}}, \citenamefont {{Jeune}}, \citenamefont
  {{Lee}}, \citenamefont {{Leon}}, \citenamefont {{Linder}}, \citenamefont
  {{Lowry}}, \citenamefont {{Matsuda}}, \citenamefont {{Matsumura}},
  \citenamefont {{Miller}}, \citenamefont {{Mizukami}}, \citenamefont
  {{Montgomery}}, \citenamefont {{Navaroli}}, \citenamefont {{Nishino}},
  \citenamefont {{Peloton}}, \citenamefont {{Poletti}}, \citenamefont
  {{Puglisi}}, \citenamefont {{Rebeiz}}, \citenamefont {{Raum}}, \citenamefont
  {{Reichardt}}, \citenamefont {{Richards}}, \citenamefont {{Ross}},
  \citenamefont {{Rotermund}}, \citenamefont {{Segawa}}, \citenamefont
  {{Sherwin}}, \citenamefont {{Shirley}}, \citenamefont {{Siritanasak}},
  \citenamefont {{Stebor}}, \citenamefont {{Stompor}}, \citenamefont
  {{Suzuki}}, \citenamefont {{Tajima}}, \citenamefont {{Takada}}, \citenamefont
  {{Takakura}}, \citenamefont {{Takatori}}, \citenamefont {{Tikhomirov}},
  \citenamefont {{Tomaru}}, \citenamefont {{Westbrook}}, \citenamefont
  {{Whitehorn}}, \citenamefont {{Yamashita}}, \citenamefont {{Zahn}},\ and\
  \citenamefont {{Zahn}}}]{polarbear2/simonsarray:2016}%
  \BibitemOpen
  \bibfield  {author} {\bibinfo {author} {\bibfnamefont {A.}~\bibnamefont
  {{Suzuki}}}, \bibinfo {author} {\bibfnamefont {P.}~\bibnamefont {{Ade}}},
  \bibinfo {author} {\bibfnamefont {Y.}~\bibnamefont {{Akiba}}}, \bibinfo
  {author} {\bibfnamefont {C.}~\bibnamefont {{Aleman}}}, \bibinfo {author}
  {\bibfnamefont {K.}~\bibnamefont {{Arnold}}}, \bibinfo {author}
  {\bibfnamefont {C.}~\bibnamefont {{Baccigalupi}}}, \bibinfo {author}
  {\bibfnamefont {B.}~\bibnamefont {{Barch}}}, \bibinfo {author} {\bibfnamefont
  {D.}~\bibnamefont {{Barron}}}, \bibinfo {author} {\bibfnamefont
  {A.}~\bibnamefont {{Bender}}}, \bibinfo {author} {\bibfnamefont
  {D.}~\bibnamefont {{Boettger}}}, \bibinfo {author} {\bibfnamefont
  {J.}~\bibnamefont {{Borrill}}}, \bibinfo {author} {\bibfnamefont
  {S.}~\bibnamefont {{Chapman}}}, \bibinfo {author} {\bibfnamefont
  {Y.}~\bibnamefont {{Chinone}}}, \bibinfo {author} {\bibfnamefont
  {A.}~\bibnamefont {{Cukierman}}}, \bibinfo {author} {\bibfnamefont
  {M.}~\bibnamefont {{Dobbs}}}, \bibinfo {author} {\bibfnamefont
  {A.}~\bibnamefont {{Ducout}}}, \bibinfo {author} {\bibfnamefont
  {R.}~\bibnamefont {{Dunner}}}, \bibinfo {author} {\bibfnamefont
  {T.}~\bibnamefont {{Elleflot}}}, \bibinfo {author} {\bibfnamefont
  {J.}~\bibnamefont {{Errard}}}, \bibinfo {author} {\bibfnamefont
  {G.}~\bibnamefont {{Fabbian}}}, \bibinfo {author} {\bibfnamefont
  {S.}~\bibnamefont {{Feeney}}}, \bibinfo {author} {\bibfnamefont
  {C.}~\bibnamefont {{Feng}}}, \bibinfo {author} {\bibfnamefont
  {T.}~\bibnamefont {{Fujino}}}, \bibinfo {author} {\bibfnamefont
  {G.}~\bibnamefont {{Fuller}}}, \bibinfo {author} {\bibfnamefont
  {A.}~\bibnamefont {{Gilbert}}}, \bibinfo {author} {\bibfnamefont
  {N.}~\bibnamefont {{Goeckner-Wald}}}, \bibinfo {author} {\bibfnamefont
  {J.}~\bibnamefont {{Groh}}}, \bibinfo {author} {\bibfnamefont {T.~D.}\
  \bibnamefont {{Haan}}}, \bibinfo {author} {\bibfnamefont {G.}~\bibnamefont
  {{Hall}}}, \bibinfo {author} {\bibfnamefont {N.}~\bibnamefont {{Halverson}}},
  \bibinfo {author} {\bibfnamefont {T.}~\bibnamefont {{Hamada}}}, \bibinfo
  {author} {\bibfnamefont {M.}~\bibnamefont {{Hasegawa}}}, \bibinfo {author}
  {\bibfnamefont {K.}~\bibnamefont {{Hattori}}}, \bibinfo {author}
  {\bibfnamefont {M.}~\bibnamefont {{Hazumi}}}, \bibinfo {author}
  {\bibfnamefont {C.}~\bibnamefont {{Hill}}}, \bibinfo {author} {\bibfnamefont
  {W.}~\bibnamefont {{Holzapfel}}}, \bibinfo {author} {\bibfnamefont
  {Y.}~\bibnamefont {{Hori}}}, \bibinfo {author} {\bibfnamefont
  {L.}~\bibnamefont {{Howe}}}, \bibinfo {author} {\bibfnamefont
  {Y.}~\bibnamefont {{Inoue}}}, \bibinfo {author} {\bibfnamefont
  {F.}~\bibnamefont {{Irie}}}, \bibinfo {author} {\bibfnamefont
  {G.}~\bibnamefont {{Jaehnig}}}, \bibinfo {author} {\bibfnamefont
  {A.}~\bibnamefont {{Jaffe}}}, \bibinfo {author} {\bibfnamefont
  {O.}~\bibnamefont {{Jeong}}}, \bibinfo {author} {\bibfnamefont
  {N.}~\bibnamefont {{Katayama}}}, \bibinfo {author} {\bibfnamefont
  {J.}~\bibnamefont {{Kaufman}}}, \bibinfo {author} {\bibfnamefont
  {K.}~\bibnamefont {{Kazemzadeh}}}, \bibinfo {author} {\bibfnamefont
  {B.}~\bibnamefont {{Keating}}}, \bibinfo {author} {\bibfnamefont
  {Z.}~\bibnamefont {{Kermish}}}, \bibinfo {author} {\bibfnamefont
  {R.}~\bibnamefont {{Keskitalo}}}, \bibinfo {author} {\bibfnamefont
  {T.}~\bibnamefont {{Kisner}}}, \bibinfo {author} {\bibfnamefont
  {A.}~\bibnamefont {{Kusaka}}}, \bibinfo {author} {\bibfnamefont {M.~L.}\
  \bibnamefont {{Jeune}}}, \bibinfo {author} {\bibfnamefont {A.}~\bibnamefont
  {{Lee}}}, \bibinfo {author} {\bibfnamefont {D.}~\bibnamefont {{Leon}}},
  \bibinfo {author} {\bibfnamefont {E.}~\bibnamefont {{Linder}}}, \bibinfo
  {author} {\bibfnamefont {L.}~\bibnamefont {{Lowry}}}, \bibinfo {author}
  {\bibfnamefont {F.}~\bibnamefont {{Matsuda}}}, \bibinfo {author}
  {\bibfnamefont {T.}~\bibnamefont {{Matsumura}}}, \bibinfo {author}
  {\bibfnamefont {N.}~\bibnamefont {{Miller}}}, \bibinfo {author}
  {\bibfnamefont {K.}~\bibnamefont {{Mizukami}}}, \bibinfo {author}
  {\bibfnamefont {J.}~\bibnamefont {{Montgomery}}}, \bibinfo {author}
  {\bibfnamefont {M.}~\bibnamefont {{Navaroli}}}, \bibinfo {author}
  {\bibfnamefont {H.}~\bibnamefont {{Nishino}}}, \bibinfo {author}
  {\bibfnamefont {J.}~\bibnamefont {{Peloton}}}, \bibinfo {author}
  {\bibfnamefont {D.}~\bibnamefont {{Poletti}}}, \bibinfo {author}
  {\bibfnamefont {G.}~\bibnamefont {{Puglisi}}}, \bibinfo {author}
  {\bibfnamefont {G.}~\bibnamefont {{Rebeiz}}}, \bibinfo {author}
  {\bibfnamefont {C.}~\bibnamefont {{Raum}}}, \bibinfo {author} {\bibfnamefont
  {C.}~\bibnamefont {{Reichardt}}}, \bibinfo {author} {\bibfnamefont
  {P.}~\bibnamefont {{Richards}}}, \bibinfo {author} {\bibfnamefont
  {C.}~\bibnamefont {{Ross}}}, \bibinfo {author} {\bibfnamefont
  {K.}~\bibnamefont {{Rotermund}}}, \bibinfo {author} {\bibfnamefont
  {Y.}~\bibnamefont {{Segawa}}}, \bibinfo {author} {\bibfnamefont
  {B.}~\bibnamefont {{Sherwin}}}, \bibinfo {author} {\bibfnamefont
  {I.}~\bibnamefont {{Shirley}}}, \bibinfo {author} {\bibfnamefont
  {P.}~\bibnamefont {{Siritanasak}}}, \bibinfo {author} {\bibfnamefont
  {N.}~\bibnamefont {{Stebor}}}, \bibinfo {author} {\bibfnamefont
  {R.}~\bibnamefont {{Stompor}}}, \bibinfo {author} {\bibfnamefont
  {J.}~\bibnamefont {{Suzuki}}}, \bibinfo {author} {\bibfnamefont
  {O.}~\bibnamefont {{Tajima}}}, \bibinfo {author} {\bibfnamefont
  {S.}~\bibnamefont {{Takada}}}, \bibinfo {author} {\bibfnamefont
  {S.}~\bibnamefont {{Takakura}}}, \bibinfo {author} {\bibfnamefont
  {S.}~\bibnamefont {{Takatori}}}, \bibinfo {author} {\bibfnamefont
  {A.}~\bibnamefont {{Tikhomirov}}}, \bibinfo {author} {\bibfnamefont
  {T.}~\bibnamefont {{Tomaru}}}, \bibinfo {author} {\bibfnamefont
  {B.}~\bibnamefont {{Westbrook}}}, \bibinfo {author} {\bibfnamefont
  {N.}~\bibnamefont {{Whitehorn}}}, \bibinfo {author} {\bibfnamefont
  {T.}~\bibnamefont {{Yamashita}}}, \bibinfo {author} {\bibfnamefont
  {A.}~\bibnamefont {{Zahn}}}, \ and\ \bibinfo {author} {\bibfnamefont
  {O.}~\bibnamefont {{Zahn}}},\ }\href {\doibase 10.1007/s10909-015-1425-4}
  {\bibfield  {journal} {\bibinfo  {journal} {Journal of Low Temperature
  Physics}\ }\textbf {\bibinfo {volume} {184}},\ \bibinfo {pages} {805}
  (\bibinfo {year} {2016})},\ \Eprint {http://arxiv.org/abs/1512.07299}
  {arXiv:1512.07299 [astro-ph.IM]} \BibitemShut {NoStop}%
\bibitem [{\citenamefont {{Chuss}}\ \emph {et~al.}(2016)\citenamefont
  {{Chuss}}, \citenamefont {{Ali}}, \citenamefont {{Amiri}}, \citenamefont
  {{Appel}}, \citenamefont {{Bennett}}, \citenamefont {{Colazo}}, \citenamefont
  {{Denis}}, \citenamefont {{D{\"u}nner}}, \citenamefont {{Essinger-Hileman}},
  \citenamefont {{Eimer}}, \citenamefont {{Fluxa}}, \citenamefont {{Gothe}},
  \citenamefont {{Halpern}}, \citenamefont {{Harrington}}, \citenamefont
  {{Hilton}}, \citenamefont {{Hinshaw}}, \citenamefont {{Hubmayr}},
  \citenamefont {{Iuliano}}, \citenamefont {{Marriage}}, \citenamefont
  {{Miller}}, \citenamefont {{Moseley}}, \citenamefont {{Mumby}}, \citenamefont
  {{Petroff}}, \citenamefont {{Reintsema}}, \citenamefont {{Rostem}},
  \citenamefont {{U-Yen}}, \citenamefont {{Watts}}, \citenamefont {{Wagner}},
  \citenamefont {{Wollack}}, \citenamefont {{Xu}},\ and\ \citenamefont
  {{Zeng}}}]{chuss/etal:2016}%
  \BibitemOpen
  \bibfield  {author} {\bibinfo {author} {\bibfnamefont {D.~T.}\ \bibnamefont
  {{Chuss}}}, \bibinfo {author} {\bibfnamefont {A.}~\bibnamefont {{Ali}}},
  \bibinfo {author} {\bibfnamefont {M.}~\bibnamefont {{Amiri}}}, \bibinfo
  {author} {\bibfnamefont {J.}~\bibnamefont {{Appel}}}, \bibinfo {author}
  {\bibfnamefont {C.~L.}\ \bibnamefont {{Bennett}}}, \bibinfo {author}
  {\bibfnamefont {F.}~\bibnamefont {{Colazo}}}, \bibinfo {author}
  {\bibfnamefont {K.~L.}\ \bibnamefont {{Denis}}}, \bibinfo {author}
  {\bibfnamefont {R.}~\bibnamefont {{D{\"u}nner}}}, \bibinfo {author}
  {\bibfnamefont {T.}~\bibnamefont {{Essinger-Hileman}}}, \bibinfo {author}
  {\bibfnamefont {J.}~\bibnamefont {{Eimer}}}, \bibinfo {author} {\bibfnamefont
  {P.}~\bibnamefont {{Fluxa}}}, \bibinfo {author} {\bibfnamefont
  {D.}~\bibnamefont {{Gothe}}}, \bibinfo {author} {\bibfnamefont
  {M.}~\bibnamefont {{Halpern}}}, \bibinfo {author} {\bibfnamefont
  {K.}~\bibnamefont {{Harrington}}}, \bibinfo {author} {\bibfnamefont
  {G.}~\bibnamefont {{Hilton}}}, \bibinfo {author} {\bibfnamefont
  {G.}~\bibnamefont {{Hinshaw}}}, \bibinfo {author} {\bibfnamefont
  {J.}~\bibnamefont {{Hubmayr}}}, \bibinfo {author} {\bibfnamefont
  {J.}~\bibnamefont {{Iuliano}}}, \bibinfo {author} {\bibfnamefont {T.~A.}\
  \bibnamefont {{Marriage}}}, \bibinfo {author} {\bibfnamefont
  {N.}~\bibnamefont {{Miller}}}, \bibinfo {author} {\bibfnamefont {S.~H.}\
  \bibnamefont {{Moseley}}}, \bibinfo {author} {\bibfnamefont {G.}~\bibnamefont
  {{Mumby}}}, \bibinfo {author} {\bibfnamefont {M.}~\bibnamefont {{Petroff}}},
  \bibinfo {author} {\bibfnamefont {C.}~\bibnamefont {{Reintsema}}}, \bibinfo
  {author} {\bibfnamefont {K.}~\bibnamefont {{Rostem}}}, \bibinfo {author}
  {\bibfnamefont {K.}~\bibnamefont {{U-Yen}}}, \bibinfo {author} {\bibfnamefont
  {D.}~\bibnamefont {{Watts}}}, \bibinfo {author} {\bibfnamefont
  {E.}~\bibnamefont {{Wagner}}}, \bibinfo {author} {\bibfnamefont {E.~J.}\
  \bibnamefont {{Wollack}}}, \bibinfo {author} {\bibfnamefont {Z.}~\bibnamefont
  {{Xu}}}, \ and\ \bibinfo {author} {\bibfnamefont {L.}~\bibnamefont
  {{Zeng}}},\ }\href {\doibase 10.1007/s10909-015-1368-9} {\bibfield  {journal}
  {\bibinfo  {journal} {Journal of Low Temperature Physics}\ }\textbf {\bibinfo
  {volume} {184}},\ \bibinfo {pages} {759} (\bibinfo {year} {2016})},\ \Eprint
  {http://arxiv.org/abs/1511.04414} {arXiv:1511.04414 [astro-ph.IM]}
  \BibitemShut {NoStop}%
\bibitem [{\citenamefont {{Krachmalnicoff}}\ \emph {et~al.}(2018)\citenamefont
  {{Krachmalnicoff}}, \citenamefont {{Carretti}}, \citenamefont
  {{Baccigalupi}}, \citenamefont {{Bernardi}}, \citenamefont {{Brown}},
  \citenamefont {{Gaensler}}, \citenamefont {{Haverkorn}}, \citenamefont
  {{Kesteven}}, \citenamefont {{Perrotta}}, \citenamefont {{Poppi}},\ and\
  \citenamefont {{Staveley-Smith}}}]{krachmalnicoff/etal:2018}%
  \BibitemOpen
  \bibfield  {author} {\bibinfo {author} {\bibfnamefont {N.}~\bibnamefont
  {{Krachmalnicoff}}}, \bibinfo {author} {\bibfnamefont {E.}~\bibnamefont
  {{Carretti}}}, \bibinfo {author} {\bibfnamefont {C.}~\bibnamefont
  {{Baccigalupi}}}, \bibinfo {author} {\bibfnamefont {G.}~\bibnamefont
  {{Bernardi}}}, \bibinfo {author} {\bibfnamefont {S.}~\bibnamefont {{Brown}}},
  \bibinfo {author} {\bibfnamefont {B.~M.}\ \bibnamefont {{Gaensler}}},
  \bibinfo {author} {\bibfnamefont {M.}~\bibnamefont {{Haverkorn}}}, \bibinfo
  {author} {\bibfnamefont {M.}~\bibnamefont {{Kesteven}}}, \bibinfo {author}
  {\bibfnamefont {F.}~\bibnamefont {{Perrotta}}}, \bibinfo {author}
  {\bibfnamefont {S.}~\bibnamefont {{Poppi}}}, \ and\ \bibinfo {author}
  {\bibfnamefont {L.}~\bibnamefont {{Staveley-Smith}}},\ }\href@noop {}
  {\bibfield  {journal} {\bibinfo  {journal} {ArXiv e-prints}\ } (\bibinfo
  {year} {2018})},\ \Eprint {http://arxiv.org/abs/1802.01145}
  {arXiv:1802.01145} \BibitemShut {NoStop}%
\bibitem [{\citenamefont {{Lewis}}\ and\ \citenamefont
  {{Challinor}}(2006)}]{lewis/challinor:2006}%
  \BibitemOpen
  \bibfield  {author} {\bibinfo {author} {\bibfnamefont {A.}~\bibnamefont
  {{Lewis}}}\ and\ \bibinfo {author} {\bibfnamefont {A.}~\bibnamefont
  {{Challinor}}},\ }\href {\doibase 10.1016/j.physrep.2006.03.002} {\bibfield
  {journal} {\bibinfo  {journal} {\physrep}\ }\textbf {\bibinfo {volume}
  {429}},\ \bibinfo {pages} {1} (\bibinfo {year} {2006})},\ \Eprint
  {http://arxiv.org/abs/astro-ph/0601594} {astro-ph/0601594} \BibitemShut
  {NoStop}%
\bibitem [{\citenamefont {{Eriksen}}\ \emph {et~al.}(2008)\citenamefont
  {{Eriksen}}, \citenamefont {{Jewell}}, \citenamefont {{Dickinson}},
  \citenamefont {{Banday}}, \citenamefont {{G{\'o}rski}},\ and\ \citenamefont
  {{Lawrence}}}]{eriksen/etal:2008}%
  \BibitemOpen
  \bibfield  {author} {\bibinfo {author} {\bibfnamefont {H.~K.}\ \bibnamefont
  {{Eriksen}}}, \bibinfo {author} {\bibfnamefont {J.~B.}\ \bibnamefont
  {{Jewell}}}, \bibinfo {author} {\bibfnamefont {C.}~\bibnamefont
  {{Dickinson}}}, \bibinfo {author} {\bibfnamefont {A.~J.}\ \bibnamefont
  {{Banday}}}, \bibinfo {author} {\bibfnamefont {K.~M.}\ \bibnamefont
  {{G{\'o}rski}}}, \ and\ \bibinfo {author} {\bibfnamefont {C.~R.}\
  \bibnamefont {{Lawrence}}},\ }\href {\doibase 10.1086/525277} {\bibfield
  {journal} {\bibinfo  {journal} {\apj}\ }\textbf {\bibinfo {volume} {676}},\
  \bibinfo {pages} {10} (\bibinfo {year} {2008})},\ \Eprint
  {http://arxiv.org/abs/0709.1058} {arXiv:0709.1058} \BibitemShut {NoStop}%
\bibitem [{\citenamefont {Alonso}\ \emph {et~al.}(2017)\citenamefont {Alonso},
  \citenamefont {Dunkley}, \citenamefont {Thorne},\ and\ \citenamefont
  {N{\ae}ss}}]{alonso17_simul_forec_primor_b_searc}%
  \BibitemOpen
  \bibfield  {author} {\bibinfo {author} {\bibfnamefont {D.}~\bibnamefont
  {Alonso}}, \bibinfo {author} {\bibfnamefont {J.}~\bibnamefont {Dunkley}},
  \bibinfo {author} {\bibfnamefont {B.}~\bibnamefont {Thorne}}, \ and\ \bibinfo
  {author} {\bibfnamefont {S.}~\bibnamefont {N{\ae}ss}},\ }\href {\doibase
  10.1103/physrevd.95.043504} {\bibfield  {journal} {\bibinfo  {journal}
  {Physical Review D}\ }\textbf {\bibinfo {volume} {95}},\ \bibinfo {pages}
  {043504} (\bibinfo {year} {2017})}\BibitemShut {NoStop}%
\bibitem [{\citenamefont {{Remazeilles}}\ \emph {et~al.}(2018)\citenamefont
  {{Remazeilles}}, \citenamefont {{Banday}}, \citenamefont {{Baccigalupi}},
  \citenamefont {{Basak}}, \citenamefont {{Bonaldi}}, \citenamefont {{De
  Zotti}}, \citenamefont {{Delabrouille}}, \citenamefont {{Dickinson}},
  \citenamefont {{Eriksen}}, \citenamefont {{Errard}}, \citenamefont
  {{Fernandez-Cobos}}, \citenamefont {{Fuskeland}}, \citenamefont
  {{Herv{\'{\i}}as-Caimapo}}, \citenamefont {{L{\'o}pez-Caniego}},
  \citenamefont {{Martinez-Gonz{\'a}lez}}, \citenamefont {{Roman}},
  \citenamefont {{Vielva}}, \citenamefont {{Wehus}}, \citenamefont
  {{Achucarro}}, \citenamefont {{Ade}}, \citenamefont {{Allison}},
  \citenamefont {{Ashdown}}, \citenamefont {{Ballardini}}, \citenamefont
  {{Banerji}}, \citenamefont {{Bartlett}}, \citenamefont {{Bartolo}},
  \citenamefont {{Baumann}}, \citenamefont {{Bersanelli}}, \citenamefont
  {{Bonato}}, \citenamefont {{Borrill}}, \citenamefont {{Bouchet}},
  \citenamefont {{Boulanger}}, \citenamefont {{Brinckmann}}, \citenamefont
  {{Bucher}}, \citenamefont {{Burigana}}, \citenamefont {{Buzzelli}},
  \citenamefont {{Cai}}, \citenamefont {{Calvo}}, \citenamefont {{Carvalho}},
  \citenamefont {{Castellano}}, \citenamefont {{Challinor}}, \citenamefont
  {{Chluba}}, \citenamefont {{Clesse}}, \citenamefont {{Colantoni}},
  \citenamefont {{Coppolecchia}}, \citenamefont {{Crook}}, \citenamefont
  {{D'Alessandro}}, \citenamefont {{de Bernardis}}, \citenamefont {{de
  Gasperis}}, \citenamefont {{Diego}}, \citenamefont {{Di Valentino}},
  \citenamefont {{Feeney}}, \citenamefont {{Ferraro}}, \citenamefont
  {{Finelli}}, \citenamefont {{Forastieri}}, \citenamefont {{Galli}},
  \citenamefont {{Genova-Santos}}, \citenamefont {{Gerbino}}, \citenamefont
  {{Gonz{\'a}lez-Nuevo}}, \citenamefont {{Grandis}}, \citenamefont
  {{Greenslade}}, \citenamefont {{Hagstotz}}, \citenamefont {{Hanany}},
  \citenamefont {{Handley}}, \citenamefont {{Hernandez-Monteagudo}},
  \citenamefont {{Hills}}, \citenamefont {{Hivon}}, \citenamefont {{Kiiveri}},
  \citenamefont {{Kisner}}, \citenamefont {{Kitching}}, \citenamefont {{Kunz}},
  \citenamefont {{Kurki-Suonio}}, \citenamefont {{Lamagna}}, \citenamefont
  {{Lasenby}}, \citenamefont {{Lattanzi}}, \citenamefont {{Lesgourgues}},
  \citenamefont {{Lewis}}, \citenamefont {{Liguori}}, \citenamefont
  {{Lindholm}}, \citenamefont {{Luzzi}}, \citenamefont {{Maffei}},
  \citenamefont {{Martins}}, \citenamefont {{Masi}}, \citenamefont
  {{Matarrese}}, \citenamefont {{McCarthy}}, \citenamefont {{Melin}},
  \citenamefont {{Melchiorri}}, \citenamefont {{Molinari}}, \citenamefont
  {{Monfardini}}, \citenamefont {{Natoli}}, \citenamefont {{Negrello}},
  \citenamefont {{Notari}}, \citenamefont {{Paiella}}, \citenamefont
  {{Paoletti}}, \citenamefont {{Patanchon}}, \citenamefont {{Piat}},
  \citenamefont {{Pisano}}, \citenamefont {{Polastri}}, \citenamefont
  {{Polenta}}, \citenamefont {{Pollo}}, \citenamefont {{Poulin}}, \citenamefont
  {{Quartin}}, \citenamefont {{Rubino-Martin}}, \citenamefont {{Salvati}},
  \citenamefont {{Tartari}}, \citenamefont {{Tomasi}}, \citenamefont
  {{Tramonte}}, \citenamefont {{Trappe}}, \citenamefont {{Trombetti}},
  \citenamefont {{Tucker}}, \citenamefont {{Valiviita}}, \citenamefont {{Van de
  Weijgaert}}, \citenamefont {{van Tent}}, \citenamefont {{Vennin}},
  \citenamefont {{Vittorio}}, \citenamefont {{Young}},\ and\ \citenamefont
  {{Zannoni}}}]{core_component_separation:2018}%
  \BibitemOpen
  \bibfield  {author} {\bibinfo {author} {\bibfnamefont {M.}~\bibnamefont
  {{Remazeilles}}}, \bibinfo {author} {\bibfnamefont {A.~J.}\ \bibnamefont
  {{Banday}}}, \bibinfo {author} {\bibfnamefont {C.}~\bibnamefont
  {{Baccigalupi}}}, \bibinfo {author} {\bibfnamefont {S.}~\bibnamefont
  {{Basak}}}, \bibinfo {author} {\bibfnamefont {A.}~\bibnamefont {{Bonaldi}}},
  \bibinfo {author} {\bibfnamefont {G.}~\bibnamefont {{De Zotti}}}, \bibinfo
  {author} {\bibfnamefont {J.}~\bibnamefont {{Delabrouille}}}, \bibinfo
  {author} {\bibfnamefont {C.}~\bibnamefont {{Dickinson}}}, \bibinfo {author}
  {\bibfnamefont {H.~K.}\ \bibnamefont {{Eriksen}}}, \bibinfo {author}
  {\bibfnamefont {J.}~\bibnamefont {{Errard}}}, \bibinfo {author}
  {\bibfnamefont {R.}~\bibnamefont {{Fernandez-Cobos}}}, \bibinfo {author}
  {\bibfnamefont {U.}~\bibnamefont {{Fuskeland}}}, \bibinfo {author}
  {\bibfnamefont {C.}~\bibnamefont {{Herv{\'{\i}}as-Caimapo}}}, \bibinfo
  {author} {\bibfnamefont {M.}~\bibnamefont {{L{\'o}pez-Caniego}}}, \bibinfo
  {author} {\bibfnamefont {E.}~\bibnamefont {{Martinez-Gonz{\'a}lez}}},
  \bibinfo {author} {\bibfnamefont {M.}~\bibnamefont {{Roman}}}, \bibinfo
  {author} {\bibfnamefont {P.}~\bibnamefont {{Vielva}}}, \bibinfo {author}
  {\bibfnamefont {I.}~\bibnamefont {{Wehus}}}, \bibinfo {author} {\bibfnamefont
  {A.}~\bibnamefont {{Achucarro}}}, \bibinfo {author} {\bibfnamefont
  {P.}~\bibnamefont {{Ade}}}, \bibinfo {author} {\bibfnamefont
  {R.}~\bibnamefont {{Allison}}}, \bibinfo {author} {\bibfnamefont
  {M.}~\bibnamefont {{Ashdown}}}, \bibinfo {author} {\bibfnamefont
  {M.}~\bibnamefont {{Ballardini}}}, \bibinfo {author} {\bibfnamefont
  {R.}~\bibnamefont {{Banerji}}}, \bibinfo {author} {\bibfnamefont
  {J.}~\bibnamefont {{Bartlett}}}, \bibinfo {author} {\bibfnamefont
  {N.}~\bibnamefont {{Bartolo}}}, \bibinfo {author} {\bibfnamefont
  {D.}~\bibnamefont {{Baumann}}}, \bibinfo {author} {\bibfnamefont
  {M.}~\bibnamefont {{Bersanelli}}}, \bibinfo {author} {\bibfnamefont
  {M.}~\bibnamefont {{Bonato}}}, \bibinfo {author} {\bibfnamefont
  {J.}~\bibnamefont {{Borrill}}}, \bibinfo {author} {\bibfnamefont
  {F.}~\bibnamefont {{Bouchet}}}, \bibinfo {author} {\bibfnamefont
  {F.}~\bibnamefont {{Boulanger}}}, \bibinfo {author} {\bibfnamefont
  {T.}~\bibnamefont {{Brinckmann}}}, \bibinfo {author} {\bibfnamefont
  {M.}~\bibnamefont {{Bucher}}}, \bibinfo {author} {\bibfnamefont
  {C.}~\bibnamefont {{Burigana}}}, \bibinfo {author} {\bibfnamefont
  {A.}~\bibnamefont {{Buzzelli}}}, \bibinfo {author} {\bibfnamefont {Z.-Y.}\
  \bibnamefont {{Cai}}}, \bibinfo {author} {\bibfnamefont {M.}~\bibnamefont
  {{Calvo}}}, \bibinfo {author} {\bibfnamefont {C.-S.}\ \bibnamefont
  {{Carvalho}}}, \bibinfo {author} {\bibfnamefont {G.}~\bibnamefont
  {{Castellano}}}, \bibinfo {author} {\bibfnamefont {A.}~\bibnamefont
  {{Challinor}}}, \bibinfo {author} {\bibfnamefont {J.}~\bibnamefont
  {{Chluba}}}, \bibinfo {author} {\bibfnamefont {S.}~\bibnamefont {{Clesse}}},
  \bibinfo {author} {\bibfnamefont {I.}~\bibnamefont {{Colantoni}}}, \bibinfo
  {author} {\bibfnamefont {A.}~\bibnamefont {{Coppolecchia}}}, \bibinfo
  {author} {\bibfnamefont {M.}~\bibnamefont {{Crook}}}, \bibinfo {author}
  {\bibfnamefont {G.}~\bibnamefont {{D'Alessandro}}}, \bibinfo {author}
  {\bibfnamefont {P.}~\bibnamefont {{de Bernardis}}}, \bibinfo {author}
  {\bibfnamefont {G.}~\bibnamefont {{de Gasperis}}}, \bibinfo {author}
  {\bibfnamefont {J.-M.}\ \bibnamefont {{Diego}}}, \bibinfo {author}
  {\bibfnamefont {E.}~\bibnamefont {{Di Valentino}}}, \bibinfo {author}
  {\bibfnamefont {S.}~\bibnamefont {{Feeney}}}, \bibinfo {author}
  {\bibfnamefont {S.}~\bibnamefont {{Ferraro}}}, \bibinfo {author}
  {\bibfnamefont {F.}~\bibnamefont {{Finelli}}}, \bibinfo {author}
  {\bibfnamefont {F.}~\bibnamefont {{Forastieri}}}, \bibinfo {author}
  {\bibfnamefont {S.}~\bibnamefont {{Galli}}}, \bibinfo {author} {\bibfnamefont
  {R.}~\bibnamefont {{Genova-Santos}}}, \bibinfo {author} {\bibfnamefont
  {M.}~\bibnamefont {{Gerbino}}}, \bibinfo {author} {\bibfnamefont
  {J.}~\bibnamefont {{Gonz{\'a}lez-Nuevo}}}, \bibinfo {author} {\bibfnamefont
  {S.}~\bibnamefont {{Grandis}}}, \bibinfo {author} {\bibfnamefont
  {J.}~\bibnamefont {{Greenslade}}}, \bibinfo {author} {\bibfnamefont
  {S.}~\bibnamefont {{Hagstotz}}}, \bibinfo {author} {\bibfnamefont
  {S.}~\bibnamefont {{Hanany}}}, \bibinfo {author} {\bibfnamefont
  {W.}~\bibnamefont {{Handley}}}, \bibinfo {author} {\bibfnamefont
  {C.}~\bibnamefont {{Hernandez-Monteagudo}}}, \bibinfo {author} {\bibfnamefont
  {M.}~\bibnamefont {{Hills}}}, \bibinfo {author} {\bibfnamefont
  {E.}~\bibnamefont {{Hivon}}}, \bibinfo {author} {\bibfnamefont
  {K.}~\bibnamefont {{Kiiveri}}}, \bibinfo {author} {\bibfnamefont
  {T.}~\bibnamefont {{Kisner}}}, \bibinfo {author} {\bibfnamefont
  {T.}~\bibnamefont {{Kitching}}}, \bibinfo {author} {\bibfnamefont
  {M.}~\bibnamefont {{Kunz}}}, \bibinfo {author} {\bibfnamefont
  {H.}~\bibnamefont {{Kurki-Suonio}}}, \bibinfo {author} {\bibfnamefont
  {L.}~\bibnamefont {{Lamagna}}}, \bibinfo {author} {\bibfnamefont
  {A.}~\bibnamefont {{Lasenby}}}, \bibinfo {author} {\bibfnamefont
  {M.}~\bibnamefont {{Lattanzi}}}, \bibinfo {author} {\bibfnamefont
  {J.}~\bibnamefont {{Lesgourgues}}}, \bibinfo {author} {\bibfnamefont
  {A.}~\bibnamefont {{Lewis}}}, \bibinfo {author} {\bibfnamefont
  {M.}~\bibnamefont {{Liguori}}}, \bibinfo {author} {\bibfnamefont
  {V.}~\bibnamefont {{Lindholm}}}, \bibinfo {author} {\bibfnamefont
  {G.}~\bibnamefont {{Luzzi}}}, \bibinfo {author} {\bibfnamefont
  {B.}~\bibnamefont {{Maffei}}}, \bibinfo {author} {\bibfnamefont
  {C.~J.~A.~P.}\ \bibnamefont {{Martins}}}, \bibinfo {author} {\bibfnamefont
  {S.}~\bibnamefont {{Masi}}}, \bibinfo {author} {\bibfnamefont
  {S.}~\bibnamefont {{Matarrese}}}, \bibinfo {author} {\bibfnamefont
  {D.}~\bibnamefont {{McCarthy}}}, \bibinfo {author} {\bibfnamefont {J.-B.}\
  \bibnamefont {{Melin}}}, \bibinfo {author} {\bibfnamefont {A.}~\bibnamefont
  {{Melchiorri}}}, \bibinfo {author} {\bibfnamefont {D.}~\bibnamefont
  {{Molinari}}}, \bibinfo {author} {\bibfnamefont {A.}~\bibnamefont
  {{Monfardini}}}, \bibinfo {author} {\bibfnamefont {P.}~\bibnamefont
  {{Natoli}}}, \bibinfo {author} {\bibfnamefont {M.}~\bibnamefont
  {{Negrello}}}, \bibinfo {author} {\bibfnamefont {A.}~\bibnamefont
  {{Notari}}}, \bibinfo {author} {\bibfnamefont {A.}~\bibnamefont {{Paiella}}},
  \bibinfo {author} {\bibfnamefont {D.}~\bibnamefont {{Paoletti}}}, \bibinfo
  {author} {\bibfnamefont {G.}~\bibnamefont {{Patanchon}}}, \bibinfo {author}
  {\bibfnamefont {M.}~\bibnamefont {{Piat}}}, \bibinfo {author} {\bibfnamefont
  {G.}~\bibnamefont {{Pisano}}}, \bibinfo {author} {\bibfnamefont
  {L.}~\bibnamefont {{Polastri}}}, \bibinfo {author} {\bibfnamefont
  {G.}~\bibnamefont {{Polenta}}}, \bibinfo {author} {\bibfnamefont
  {A.}~\bibnamefont {{Pollo}}}, \bibinfo {author} {\bibfnamefont
  {V.}~\bibnamefont {{Poulin}}}, \bibinfo {author} {\bibfnamefont
  {M.}~\bibnamefont {{Quartin}}}, \bibinfo {author} {\bibfnamefont {J.-A.}\
  \bibnamefont {{Rubino-Martin}}}, \bibinfo {author} {\bibfnamefont
  {L.}~\bibnamefont {{Salvati}}}, \bibinfo {author} {\bibfnamefont
  {A.}~\bibnamefont {{Tartari}}}, \bibinfo {author} {\bibfnamefont
  {M.}~\bibnamefont {{Tomasi}}}, \bibinfo {author} {\bibfnamefont
  {D.}~\bibnamefont {{Tramonte}}}, \bibinfo {author} {\bibfnamefont
  {N.}~\bibnamefont {{Trappe}}}, \bibinfo {author} {\bibfnamefont
  {T.}~\bibnamefont {{Trombetti}}}, \bibinfo {author} {\bibfnamefont
  {C.}~\bibnamefont {{Tucker}}}, \bibinfo {author} {\bibfnamefont
  {J.}~\bibnamefont {{Valiviita}}}, \bibinfo {author} {\bibfnamefont
  {R.}~\bibnamefont {{Van de Weijgaert}}}, \bibinfo {author} {\bibfnamefont
  {B.}~\bibnamefont {{van Tent}}}, \bibinfo {author} {\bibfnamefont
  {V.}~\bibnamefont {{Vennin}}}, \bibinfo {author} {\bibfnamefont
  {N.}~\bibnamefont {{Vittorio}}}, \bibinfo {author} {\bibfnamefont
  {K.}~\bibnamefont {{Young}}}, \ and\ \bibinfo {author} {\bibfnamefont
  {M.}~\bibnamefont {{Zannoni}}},\ }\href {\doibase
  10.1088/1475-7516/2018/04/023} {\bibfield  {journal} {\bibinfo  {journal}
  {\jcap}\ }\textbf {\bibinfo {volume} {4}},\ \bibinfo {eid} {023} (\bibinfo
  {year} {2018})},\ \Eprint {http://arxiv.org/abs/1704.04501}
  {arXiv:1704.04501} \BibitemShut {NoStop}%
\bibitem [{Note1()}]{Note1}%
  \BibitemOpen
  \bibinfo {note} {\protect \url
  {https://github.com/bthorne93/PySM_public}}\BibitemShut {NoStop}%
\bibitem [{\citenamefont {Thorne}\ \emph {et~al.}(2017)\citenamefont {Thorne},
  \citenamefont {Dunkley}, \citenamefont {Alonso},\ and\ \citenamefont
  {N{\ae}ss}}]{thorne17_python_sky_model}%
  \BibitemOpen
  \bibfield  {author} {\bibinfo {author} {\bibfnamefont {B.}~\bibnamefont
  {Thorne}}, \bibinfo {author} {\bibfnamefont {J.}~\bibnamefont {Dunkley}},
  \bibinfo {author} {\bibfnamefont {D.}~\bibnamefont {Alonso}}, \ and\ \bibinfo
  {author} {\bibfnamefont {S.}~\bibnamefont {N{\ae}ss}},\ }\href {\doibase
  10.1093/mnras/stx949} {\bibfield  {journal} {\bibinfo  {journal} {Monthly
  Notices of the Royal Astronomical Society}\ }\textbf {\bibinfo {volume}
  {469}},\ \bibinfo {pages} {2821} (\bibinfo {year} {2017})}\BibitemShut
  {NoStop}%
\bibitem [{\citenamefont {Bennett}\ \emph {et~al.}(2013)\citenamefont
  {Bennett}, \citenamefont {Larson}, \citenamefont {Weiland}, \citenamefont
  {Jarosik}, \citenamefont {Hinshaw}, \citenamefont {Odegard}, \citenamefont
  {Smith}, \citenamefont {Hill}, \citenamefont {Gold}, \citenamefont {Halpern},
  \citenamefont {Komatsu}, \citenamefont {Nolta}, \citenamefont {Page},
  \citenamefont {Spergel}, \citenamefont {Wollack}, \citenamefont {Dunkley},
  \citenamefont {Kogut}, \citenamefont {Limon}, \citenamefont {Meyer},
  \citenamefont {Tucker},\ and\ \citenamefont
  {Wright}}]{bennett13_nine_yearw_microw_anisot_probe_wmap_obser}%
  \BibitemOpen
  \bibfield  {author} {\bibinfo {author} {\bibfnamefont {C.~L.}\ \bibnamefont
  {Bennett}}, \bibinfo {author} {\bibfnamefont {D.}~\bibnamefont {Larson}},
  \bibinfo {author} {\bibfnamefont {J.~L.}\ \bibnamefont {Weiland}}, \bibinfo
  {author} {\bibfnamefont {N.}~\bibnamefont {Jarosik}}, \bibinfo {author}
  {\bibfnamefont {G.}~\bibnamefont {Hinshaw}}, \bibinfo {author} {\bibfnamefont
  {N.}~\bibnamefont {Odegard}}, \bibinfo {author} {\bibfnamefont {K.~M.}\
  \bibnamefont {Smith}}, \bibinfo {author} {\bibfnamefont {R.~S.}\ \bibnamefont
  {Hill}}, \bibinfo {author} {\bibfnamefont {B.}~\bibnamefont {Gold}}, \bibinfo
  {author} {\bibfnamefont {M.}~\bibnamefont {Halpern}}, \bibinfo {author}
  {\bibfnamefont {E.}~\bibnamefont {Komatsu}}, \bibinfo {author} {\bibfnamefont
  {M.~R.}\ \bibnamefont {Nolta}}, \bibinfo {author} {\bibfnamefont
  {L.}~\bibnamefont {Page}}, \bibinfo {author} {\bibfnamefont {D.~N.}\
  \bibnamefont {Spergel}}, \bibinfo {author} {\bibfnamefont {E.}~\bibnamefont
  {Wollack}}, \bibinfo {author} {\bibfnamefont {J.}~\bibnamefont {Dunkley}},
  \bibinfo {author} {\bibfnamefont {A.}~\bibnamefont {Kogut}}, \bibinfo
  {author} {\bibfnamefont {M.}~\bibnamefont {Limon}}, \bibinfo {author}
  {\bibfnamefont {S.~S.}\ \bibnamefont {Meyer}}, \bibinfo {author}
  {\bibfnamefont {G.~S.}\ \bibnamefont {Tucker}}, \ and\ \bibinfo {author}
  {\bibfnamefont {E.~L.}\ \bibnamefont {Wright}},\ }\href {\doibase
  10.1088/0067-0049/208/2/20} {\bibfield  {journal} {\bibinfo  {journal} {The
  Astrophysical Journal Supplement Series}\ }\textbf {\bibinfo {volume}
  {208}},\ \bibinfo {pages} {20} (\bibinfo {year} {2013})}\BibitemShut
  {NoStop}%
\bibitem [{Note2()}]{Note2}%
  \BibitemOpen
  \bibinfo {note} {\protect \url
  {https://github.com/amaurea/taylens}}\BibitemShut {NoStop}%
\bibitem [{\citenamefont {{Naess}}\ and\ \citenamefont
  {{Louis}}(2013)}]{naess/louis:2013}%
  \BibitemOpen
  \bibfield  {author} {\bibinfo {author} {\bibfnamefont {S.~K.}\ \bibnamefont
  {{Naess}}}\ and\ \bibinfo {author} {\bibfnamefont {T.}~\bibnamefont
  {{Louis}}},\ }\href {\doibase 10.1088/1475-7516/2013/09/001} {\bibfield
  {journal} {\bibinfo  {journal} {\jcap}\ }\textbf {\bibinfo {volume} {9}},\
  \bibinfo {eid} {001} (\bibinfo {year} {2013})},\ \Eprint
  {http://arxiv.org/abs/1307.0719} {arXiv:1307.0719} \BibitemShut {NoStop}%
\bibitem [{\citenamefont {{Stevens}}\ \emph {et~al.}(2018)\citenamefont
  {{Stevens}}, \citenamefont {{Goeckner-Wald}}, \citenamefont {{Keskitalo}},
  \citenamefont {{McCallum}}, \citenamefont {{Ali}}, \citenamefont {{Borrill}},
  \citenamefont {{Brown}}, \citenamefont {{Chinone}}, \citenamefont
  {{Gallardo}}, \citenamefont {{Kusaka}}, \citenamefont {{Lee}}, \citenamefont
  {{McMahon}}, \citenamefont {{Niemack}}, \citenamefont {{Page}}, \citenamefont
  {{Puglisi}}, \citenamefont {{Salatino}}, \citenamefont {{Mak}}, \citenamefont
  {{Teply}}, \citenamefont {{Thomas}}, \citenamefont {{Vavagiakis}},
  \citenamefont {{Wollack}}, \citenamefont {{Xu}},\ and\ \citenamefont
  {{Zhu}}}]{stevens/etal:2018}%
  \BibitemOpen
  \bibfield  {author} {\bibinfo {author} {\bibfnamefont {J.~R.}\ \bibnamefont
  {{Stevens}}}, \bibinfo {author} {\bibfnamefont {N.}~\bibnamefont
  {{Goeckner-Wald}}}, \bibinfo {author} {\bibfnamefont {R.}~\bibnamefont
  {{Keskitalo}}}, \bibinfo {author} {\bibfnamefont {N.}~\bibnamefont
  {{McCallum}}}, \bibinfo {author} {\bibfnamefont {A.}~\bibnamefont {{Ali}}},
  \bibinfo {author} {\bibfnamefont {J.}~\bibnamefont {{Borrill}}}, \bibinfo
  {author} {\bibfnamefont {M.~L.}\ \bibnamefont {{Brown}}}, \bibinfo {author}
  {\bibfnamefont {Y.}~\bibnamefont {{Chinone}}}, \bibinfo {author}
  {\bibfnamefont {P.~A.}\ \bibnamefont {{Gallardo}}}, \bibinfo {author}
  {\bibfnamefont {A.}~\bibnamefont {{Kusaka}}}, \bibinfo {author}
  {\bibfnamefont {A.~T.}\ \bibnamefont {{Lee}}}, \bibinfo {author}
  {\bibfnamefont {J.}~\bibnamefont {{McMahon}}}, \bibinfo {author}
  {\bibfnamefont {M.~D.}\ \bibnamefont {{Niemack}}}, \bibinfo {author}
  {\bibfnamefont {L.}~\bibnamefont {{Page}}}, \bibinfo {author} {\bibfnamefont
  {G.}~\bibnamefont {{Puglisi}}}, \bibinfo {author} {\bibfnamefont
  {M.}~\bibnamefont {{Salatino}}}, \bibinfo {author} {\bibfnamefont {S.~Y.~D.}\
  \bibnamefont {{Mak}}}, \bibinfo {author} {\bibfnamefont {G.}~\bibnamefont
  {{Teply}}}, \bibinfo {author} {\bibfnamefont {D.~B.}\ \bibnamefont
  {{Thomas}}}, \bibinfo {author} {\bibfnamefont {E.~M.}\ \bibnamefont
  {{Vavagiakis}}}, \bibinfo {author} {\bibfnamefont {E.~J.}\ \bibnamefont
  {{Wollack}}}, \bibinfo {author} {\bibfnamefont {Z.}~\bibnamefont {{Xu}}}, \
  and\ \bibinfo {author} {\bibfnamefont {N.}~\bibnamefont {{Zhu}}},\ }in\ \href
  {\doibase 10.1117/12.2313898} {\emph {\bibinfo {booktitle} {Millimeter,
  Submillimeter, and Far-Infrared Detectors and Instrumentation for Astronomy
  IX}}},\ \bibinfo {series} {Society of Photo-Optical Instrumentation Engineers
  (SPIE) Conference Series}, Vol.\ \bibinfo {volume} {10708}\ (\bibinfo {year}
  {2018})\ p.\ \bibinfo {pages} {1070841},\ \Eprint
  {http://arxiv.org/abs/1808.05131} {arXiv:1808.05131 [astro-ph.IM]}
  \BibitemShut {NoStop}%
\bibitem [{\citenamefont {{Kusaka}}\ \emph {et~al.}(2018)\citenamefont
  {{Kusaka}}, \citenamefont {{Appel}}, \citenamefont {{Essinger-Hileman}},
  \citenamefont {{Beall}}, \citenamefont {{Campusano}}, \citenamefont {{Cho}},
  \citenamefont {{Choi}}, \citenamefont {{Crowley}}, \citenamefont {{Fowler}},
  \citenamefont {{Gallardo}}, \citenamefont {{Hasselfield}}, \citenamefont
  {{Hilton}}, \citenamefont {{Ho}}, \citenamefont {{Irwin}}, \citenamefont
  {{Jarosik}}, \citenamefont {{Niemack}}, \citenamefont {{Nixon}},
  \citenamefont {{Nolta}}, \citenamefont {{Page}}, \citenamefont {{Palma}},
  \citenamefont {{Parker}}, \citenamefont {{Raghunathan}}, \citenamefont
  {{Reintsema}}, \citenamefont {{Sievers}}, \citenamefont {{Simon}},
  \citenamefont {{Staggs}}, \citenamefont {{Visnjic}},\ and\ \citenamefont
  {{Yoon}}}]{kusaks/etal:2018}%
  \BibitemOpen
  \bibfield  {author} {\bibinfo {author} {\bibfnamefont {A.}~\bibnamefont
  {{Kusaka}}}, \bibinfo {author} {\bibfnamefont {J.}~\bibnamefont {{Appel}}},
  \bibinfo {author} {\bibfnamefont {T.}~\bibnamefont {{Essinger-Hileman}}},
  \bibinfo {author} {\bibfnamefont {J.~A.}\ \bibnamefont {{Beall}}}, \bibinfo
  {author} {\bibfnamefont {L.~E.}\ \bibnamefont {{Campusano}}}, \bibinfo
  {author} {\bibfnamefont {H.-M.}\ \bibnamefont {{Cho}}}, \bibinfo {author}
  {\bibfnamefont {S.~K.}\ \bibnamefont {{Choi}}}, \bibinfo {author}
  {\bibfnamefont {K.}~\bibnamefont {{Crowley}}}, \bibinfo {author}
  {\bibfnamefont {J.~W.}\ \bibnamefont {{Fowler}}}, \bibinfo {author}
  {\bibfnamefont {P.}~\bibnamefont {{Gallardo}}}, \bibinfo {author}
  {\bibfnamefont {M.}~\bibnamefont {{Hasselfield}}}, \bibinfo {author}
  {\bibfnamefont {G.}~\bibnamefont {{Hilton}}}, \bibinfo {author}
  {\bibfnamefont {S.-P.~P.}\ \bibnamefont {{Ho}}}, \bibinfo {author}
  {\bibfnamefont {K.}~\bibnamefont {{Irwin}}}, \bibinfo {author} {\bibfnamefont
  {N.}~\bibnamefont {{Jarosik}}}, \bibinfo {author} {\bibfnamefont {M.~D.}\
  \bibnamefont {{Niemack}}}, \bibinfo {author} {\bibfnamefont {G.~W.}\
  \bibnamefont {{Nixon}}}, \bibinfo {author} {\bibfnamefont {M.}~\bibnamefont
  {{Nolta}}}, \bibinfo {author} {\bibfnamefont {J.}~\bibnamefont {{Page}},
  \bibfnamefont {Lyman~A.}}, \bibinfo {author} {\bibfnamefont {G.~A.}\
  \bibnamefont {{Palma}}}, \bibinfo {author} {\bibfnamefont {L.}~\bibnamefont
  {{Parker}}}, \bibinfo {author} {\bibfnamefont {S.}~\bibnamefont
  {{Raghunathan}}}, \bibinfo {author} {\bibfnamefont {C.~D.}\ \bibnamefont
  {{Reintsema}}}, \bibinfo {author} {\bibfnamefont {J.}~\bibnamefont
  {{Sievers}}}, \bibinfo {author} {\bibfnamefont {S.~M.}\ \bibnamefont
  {{Simon}}}, \bibinfo {author} {\bibfnamefont {S.~T.}\ \bibnamefont
  {{Staggs}}}, \bibinfo {author} {\bibfnamefont {K.}~\bibnamefont {{Visnjic}}},
  \ and\ \bibinfo {author} {\bibfnamefont {K.-W.}\ \bibnamefont {{Yoon}}},\
  }\href {\doibase 10.1088/1475-7516/2018/09/005} {\bibfield  {journal}
  {\bibinfo  {journal} {Journal of Cosmology and Astro-Particle Physics}\
  }\textbf {\bibinfo {volume} {2018}},\ \bibinfo {eid} {005} (\bibinfo {year}
  {2018})},\ \Eprint {http://arxiv.org/abs/1801.01218} {arXiv:1801.01218
  [astro-ph.CO]} \BibitemShut {NoStop}%
\bibitem [{\citenamefont {{BICEP2 Collaboration}}\ and\ \citenamefont {{Keck
  Array Collaboration}}(2016)}]{bk:2016}%
  \BibitemOpen
  \bibfield  {author} {\bibinfo {author} {\bibnamefont {{BICEP2
  Collaboration}}}\ and\ \bibinfo {author} {\bibnamefont {{Keck Array
  Collaboration}}},\ }\href {\doibase 10.1103/PhysRevLett.116.031302}
  {\bibfield  {journal} {\bibinfo  {journal} {\prl}\ }\textbf {\bibinfo
  {volume} {116}},\ \bibinfo {eid} {031302} (\bibinfo {year} {2016})},\ \Eprint
  {http://arxiv.org/abs/1510.09217} {arXiv:1510.09217 [astro-ph.CO]}
  \BibitemShut {NoStop}%
\bibitem [{\citenamefont {{QUIET Collaboration}}(2011)}]{quiet:2011}%
  \BibitemOpen
  \bibfield  {author} {\bibinfo {author} {\bibnamefont {{QUIET
  Collaboration}}},\ }\href {\doibase 10.1088/0004-637X/741/2/111} {\bibfield
  {journal} {\bibinfo  {journal} {\apj}\ }\textbf {\bibinfo {volume} {741}},\
  \bibinfo {eid} {111} (\bibinfo {year} {2011})},\ \Eprint
  {http://arxiv.org/abs/1012.3191} {arXiv:1012.3191 [astro-ph.CO]} \BibitemShut
  {NoStop}%
\bibitem [{\citenamefont {{QUIET Collaboration}}(2012)}]{quiet:2012}%
  \BibitemOpen
  \bibfield  {author} {\bibinfo {author} {\bibnamefont {{QUIET
  Collaboration}}},\ }\href {\doibase 10.1088/0004-637X/760/2/145} {\bibfield
  {journal} {\bibinfo  {journal} {\apj}\ }\textbf {\bibinfo {volume} {760}},\
  \bibinfo {eid} {145} (\bibinfo {year} {2012})},\ \Eprint
  {http://arxiv.org/abs/1207.5034} {arXiv:1207.5034 [astro-ph.CO]} \BibitemShut
  {NoStop}%
\bibitem [{\citenamefont {{Grain}}\ \emph {et~al.}(2009)\citenamefont
  {{Grain}}, \citenamefont {{Tristram}},\ and\ \citenamefont
  {{Stompor}}}]{grain/etal:2009}%
  \BibitemOpen
  \bibfield  {author} {\bibinfo {author} {\bibfnamefont {J.}~\bibnamefont
  {{Grain}}}, \bibinfo {author} {\bibfnamefont {M.}~\bibnamefont {{Tristram}}},
  \ and\ \bibinfo {author} {\bibfnamefont {R.}~\bibnamefont {{Stompor}}},\
  }\href {\doibase 10.1103/PhysRevD.79.123515} {\bibfield  {journal} {\bibinfo
  {journal} {Physical Review D}\ }\textbf {\bibinfo {volume} {79}},\ \bibinfo
  {eid} {123515} (\bibinfo {year} {2009})},\ \Eprint
  {http://arxiv.org/abs/0903.2350} {arXiv:0903.2350 [astro-ph.CO]} \BibitemShut
  {NoStop}%
\bibitem [{\citenamefont {Wandelt}\ \emph {et~al.}(2001)\citenamefont
  {Wandelt}, \citenamefont {Hivon},\ and\ \citenamefont
  {G\'orski}}]{wandelt/etal:2001}%
  \BibitemOpen
  \bibfield  {author} {\bibinfo {author} {\bibfnamefont {B.~D.}\ \bibnamefont
  {Wandelt}}, \bibinfo {author} {\bibfnamefont {E.}~\bibnamefont {Hivon}}, \
  and\ \bibinfo {author} {\bibfnamefont {K.~M.}\ \bibnamefont {G\'orski}},\
  }\href {\doibase 10.1103/PhysRevD.64.083003} {\bibfield  {journal} {\bibinfo
  {journal} {Phys. Rev. D}\ }\textbf {\bibinfo {volume} {64}},\ \bibinfo
  {pages} {083003} (\bibinfo {year} {2001})}\BibitemShut {NoStop}%
\bibitem [{\citenamefont {{Challinor}}\ and\ \citenamefont
  {{Chon}}(2005)}]{challinor/chon:2005}%
  \BibitemOpen
  \bibfield  {author} {\bibinfo {author} {\bibfnamefont {A.}~\bibnamefont
  {{Challinor}}}\ and\ \bibinfo {author} {\bibfnamefont {G.}~\bibnamefont
  {{Chon}}},\ }\href {\doibase 10.1111/j.1365-2966.2005.09076.x} {\bibfield
  {journal} {\bibinfo  {journal} {MNRAS}\ }\textbf {\bibinfo {volume} {360}},\
  \bibinfo {pages} {509} (\bibinfo {year} {2005})},\ \Eprint
  {http://arxiv.org/abs/astro-ph/0410097} {astro-ph/0410097} \BibitemShut
  {NoStop}%
\bibitem [{\citenamefont {{Smith}}(2006)}]{smith:2006}%
  \BibitemOpen
  \bibfield  {author} {\bibinfo {author} {\bibfnamefont {K.~M.}\ \bibnamefont
  {{Smith}}},\ }\href {\doibase 10.1103/PhysRevD.74.083002} {\bibfield
  {journal} {\bibinfo  {journal} {\prd}\ }\textbf {\bibinfo {volume} {74}},\
  \bibinfo {eid} {083002} (\bibinfo {year} {2006})},\ \Eprint
  {http://arxiv.org/abs/astro-ph/0511629} {astro-ph/0511629} \BibitemShut
  {NoStop}%
\bibitem [{\citenamefont {{Smith}}\ and\ \citenamefont
  {{Zaldarriaga}}(2007)}]{smith/zaldarriaga:2007}%
  \BibitemOpen
  \bibfield  {author} {\bibinfo {author} {\bibfnamefont {K.~M.}\ \bibnamefont
  {{Smith}}}\ and\ \bibinfo {author} {\bibfnamefont {M.}~\bibnamefont
  {{Zaldarriaga}}},\ }\href {\doibase 10.1103/PhysRevD.76.043001} {\bibfield
  {journal} {\bibinfo  {journal} {\prd}\ }\textbf {\bibinfo {volume} {76}},\
  \bibinfo {eid} {043001} (\bibinfo {year} {2007})},\ \Eprint
  {http://arxiv.org/abs/astro-ph/0610059} {astro-ph/0610059} \BibitemShut
  {NoStop}%
\bibitem [{Note3()}]{Note3}%
  \BibitemOpen
  \bibinfo {note} {\protect \url
  {https://github.com/LSSTDESC/namaster}}\BibitemShut {NoStop}%
\bibitem [{\citenamefont {{Alonso}}\ \emph {et~al.}(2019)\citenamefont
  {{Alonso}}, \citenamefont {{Sanchez}},\ and\ \citenamefont
  {{Slosar}}}]{alonso/etal:2019}%
  \BibitemOpen
  \bibfield  {author} {\bibinfo {author} {\bibfnamefont {D.}~\bibnamefont
  {{Alonso}}}, \bibinfo {author} {\bibfnamefont {J.}~\bibnamefont {{Sanchez}}},
  \ and\ \bibinfo {author} {\bibfnamefont {A.}~\bibnamefont {{Slosar}}},\
  }\href {\doibase 10.1093/mnras/stz093} {\bibfield  {journal} {\bibinfo
  {journal} {\mnras}\ }\textbf {\bibinfo {volume} {484}},\ \bibinfo {pages}
  {4127} (\bibinfo {year} {2019})},\ \Eprint {http://arxiv.org/abs/1809.09603}
  {arXiv:1809.09603 [astro-ph.CO]} \BibitemShut {NoStop}%
\bibitem [{\citenamefont {{Hamimeche}}\ and\ \citenamefont
  {{Lewis}}(2008)}]{hamimeche/lewis:2008;}%
  \BibitemOpen
  \bibfield  {author} {\bibinfo {author} {\bibfnamefont {S.}~\bibnamefont
  {{Hamimeche}}}\ and\ \bibinfo {author} {\bibfnamefont {A.}~\bibnamefont
  {{Lewis}}},\ }\href {\doibase 10.1103/PhysRevD.77.103013} {\bibfield
  {journal} {\bibinfo  {journal} {\prd}\ }\textbf {\bibinfo {volume} {77}},\
  \bibinfo {eid} {103013} (\bibinfo {year} {2008})},\ \Eprint
  {http://arxiv.org/abs/0801.0554} {arXiv:0801.0554 [astro-ph]} \BibitemShut
  {NoStop}%
\bibitem [{\citenamefont {{Errard}}\ and\ \citenamefont
  {{Stompor}}(2019)}]{errard/stompor:2018}%
  \BibitemOpen
  \bibfield  {author} {\bibinfo {author} {\bibfnamefont {J.}~\bibnamefont
  {{Errard}}}\ and\ \bibinfo {author} {\bibfnamefont {R.}~\bibnamefont
  {{Stompor}}},\ }\href {\doibase 10.1103/PhysRevD.99.043529} {\bibfield
  {journal} {\bibinfo  {journal} {\prd}\ }\textbf {\bibinfo {volume} {99}},\
  \bibinfo {eid} {043529} (\bibinfo {year} {2019})},\ \Eprint
  {http://arxiv.org/abs/1811.00479} {arXiv:1811.00479 [astro-ph.CO]}
  \BibitemShut {NoStop}%
\bibitem [{Note4()}]{Note4}%
  \BibitemOpen
  \bibinfo {note} {\protect \href
  {http://pla.esac.esa.int/pla/aio/product-action?MAP.MAP_ID=HFI_Mask_GalPlane-apo0_2048_R2.00.fits}{Filename:
  {\protect \tt HFI\protect \_Mask\protect \_GalPlane-apo0\protect
  \_2048\protect \_R2.00.fits}}}\BibitemShut {NoStop}%
\bibitem [{\citenamefont {{Clark}}\ \emph {et~al.}(2015)\citenamefont
  {{Clark}}, \citenamefont {{Hill}}, \citenamefont {{Peek}}, \citenamefont
  {{Putman}},\ and\ \citenamefont {{Babler}}}]{clark/etal:2015}%
  \BibitemOpen
  \bibfield  {author} {\bibinfo {author} {\bibfnamefont {S.~E.}\ \bibnamefont
  {{Clark}}}, \bibinfo {author} {\bibfnamefont {J.~C.}\ \bibnamefont {{Hill}}},
  \bibinfo {author} {\bibfnamefont {J.~E.~G.}\ \bibnamefont {{Peek}}}, \bibinfo
  {author} {\bibfnamefont {M.~E.}\ \bibnamefont {{Putman}}}, \ and\ \bibinfo
  {author} {\bibfnamefont {B.~L.}\ \bibnamefont {{Babler}}},\ }\href {\doibase
  10.1103/PhysRevLett.115.241302} {\bibfield  {journal} {\bibinfo  {journal}
  {\prl}\ }\textbf {\bibinfo {volume} {115}},\ \bibinfo {eid} {241302}
  (\bibinfo {year} {2015})},\ \Eprint {http://arxiv.org/abs/1508.07005}
  {arXiv:1508.07005 [astro-ph.CO]} \BibitemShut {NoStop}%
\bibitem [{\citenamefont {{Clark}}(2018)}]{clark:2018}%
  \BibitemOpen
  \bibfield  {author} {\bibinfo {author} {\bibfnamefont {S.~E.}\ \bibnamefont
  {{Clark}}},\ }\href {\doibase 10.3847/2041-8213/aabb54} {\bibfield  {journal}
  {\bibinfo  {journal} {\apj}\ }\textbf {\bibinfo {volume} {857}},\ \bibinfo
  {eid} {L10} (\bibinfo {year} {2018})},\ \Eprint
  {http://arxiv.org/abs/1802.00011} {arXiv:1802.00011 [astro-ph.GA]}
  \BibitemShut {NoStop}%
\bibitem [{\citenamefont {{Chluba}}\ \emph {et~al.}(2017)\citenamefont
  {{Chluba}}, \citenamefont {{Hill}},\ and\ \citenamefont
  {{Abitbol}}}]{chluba/etal:2017}%
  \BibitemOpen
  \bibfield  {author} {\bibinfo {author} {\bibfnamefont {J.}~\bibnamefont
  {{Chluba}}}, \bibinfo {author} {\bibfnamefont {J.~C.}\ \bibnamefont
  {{Hill}}}, \ and\ \bibinfo {author} {\bibfnamefont {M.~H.}\ \bibnamefont
  {{Abitbol}}},\ }\href {\doibase 10.1093/mnras/stx1982} {\bibfield  {journal}
  {\bibinfo  {journal} {\mnras}\ }\textbf {\bibinfo {volume} {472}},\ \bibinfo
  {pages} {1195} (\bibinfo {year} {2017})},\ \Eprint
  {http://arxiv.org/abs/1701.00274} {arXiv:1701.00274 [astro-ph.CO]}
  \BibitemShut {NoStop}%
\bibitem [{Note5()}]{Note5}%
  \BibitemOpen
  \bibinfo {note} {\protect \url {www.bxb.space}}\BibitemShut {NoStop}%
\end{thebibliography}%
\end{document}